\documentclass[final,3p,times]{elsarticle}

\usepackage{amsmath,amssymb,amsfonts}
\usepackage{algorithmic}
\usepackage{algorithm}
\usepackage{graphicx}
\usepackage{booktabs}
\usepackage{tablefootnote}
\usepackage{threeparttable}
\usepackage{subfigure}
\usepackage{enumitem}
\usepackage{xcolor}
\usepackage{xspace}
\usepackage{multirow}
\usepackage{array}
\usepackage{tabularx}
\usepackage{adjustbox}
\newcommand{\method}{SST-DPN\xspace}

\begin{document}

\begin{frontmatter}

\title{A Spatial-Spectral and Temporal Dual Prototype Network for Motor Imagery Brain-Computer Interface}

\author[label1]{Can Han}
\ead{hancan@sjtu.edu.cn}
\author[label1]{Chen Liu} 
\ead{lchen1206@sjtu.edu.cn}
\author[label2]{Yaqi Wang} 
\ead{wangyaqi@cuz.edu.cn}
\author[label1]{Crystal Cai} 
\ead{crystal.cai@sjtu.edu.cn}
\author[label3]{Jun Wang\corref{corresponding}}
\ead{wangjun@hzcu.edu.cn}
\author[label1]{Dahong Qian\corref{corresponding}}
\ead{dahong.qian@sjtu.edu.cn}
\cortext[corresponding]{Corresponding author.}

\affiliation[label1]{organization={School of Biomedical Engineering, Shanghai Jiao Tong University},
            city={Shanghai},
            postcode={200240}, 
            country={China}}
\affiliation[label2]{organization={College of Media Engineering, Communication University of Zhejiang},
            city={Hangzhou},
            postcode={310018}, 
            country={China}}
\affiliation[label3]{organization={School of Computer and Computing Science, Hang Zhou City University},
            city={Hangzhou},
            postcode={310015}, 
            country={China}}

\begin{abstract}
Motor imagery electroencephalogram (MI-EEG) decoding plays a crucial role in developing motor imagery brain-computer interfaces (MI-BCIs). 
However, decoding intentions from MI remains challenging due to the inherent complexity of EEG signals relative to the small-sample size.
To address this issue, we propose a spatial-spectral and temporal dual prototype network (SST-DPN).
First, we design a lightweight attention mechanism to uniformly model the spatial-spectral relationships across multiple EEG electrodes, enabling the extraction of powerful spatial-spectral features. Then, we develop a multi-scale variance pooling module tailored for EEG signals to capture long-term temporal features. This module is parameter-free and computationally efficient, offering clear advantages over the widely used transformer models.
Furthermore, we introduce dual prototype learning to optimize the feature space distribution and training process, thereby improving the model's generalization ability on small-sample MI datasets. 
Our experimental results show that the \method outperforms state-of-the-art models with superior classification accuracy (84.11\% for dataset BCI4-2A, 86.65\% for dataset BCI4-2B).
Additionally, we use the BCI3-4A dataset with fewer training data to further validate the generalization ability of the proposed \method. We also achieve superior performance with 82.03\% classification accuracy.
Benefiting from the lightweight parameters and superior decoding accuracy, our \method shows great potential for practical MI-BCI applications.
The code is publicly available at https://github.com/hancan16/\method.
\end{abstract}


\end{frontmatter}

\section{Introduction}
\label{sec:introduction}
Brain-computer interface (BCI) systems enable non-muscular communication between users and machines by interpreting users’ neural activity patterns~\cite{gao2003bci}. 
In BCI applications, electroencephalogram (EEG) has become increasingly popular due to its non-invasive nature and cost-effectiveness.
Motor Imagery (MI)~\cite{abiri2019comprehensive} is the mental rehearsal of movement execution without any physical movement.
When participants visualize moving parts of their body, specific areas of the brain experience energy changes known as event-related desynchronization/synchronization (ERD/ERS). These changes can be recorded via EEG and used to discriminate motor intent~\cite{baniqued2021brain}.
The MI-BCIs have garnered significant attention for their ability to decode user motor intentions from EEG signals. It has been successfully applied in various fields, such as stroke rehabilitation~\cite{ang2016eeg}, wheelchair control~\cite{long2012hybrid}, and exoskeleton robot arm control~\cite{edelman2019noninvasive}.

Advancements in deep learning (DL) have significantly increased the accuracy of decoding EEG signals for MI-based BCI applications~\cite{schirrmeister2017deep,lawhern2018eegnet,liu2022fbmsnet}, yet several issues still hinder DL models from reaching practical use~\cite{gu2023beyond}.
EEG signals are often affected by noise and artifacts, leading to a low signal-to-noise ratio (SNR). In addition, the complex spatial-spectral coupling and high temporal variability of these signals make decoding MI-EEG data particularly challenging~\cite{zhang2021survey}. Thus, effectively extracting discriminative features from EEG data is both difficult and essential.
At the same time, EEG data collection is limited by a shortage of training samples, due to labor-intensive calibration, variability in participants' responses, and data privacy concerns~\cite{li2022meta,han2023noise}. This scarcity of data increases the risk of model overfitting, ultimately weakening its ability to generalize to new test data.
To this end, this paper encompasses the following three key aspects of EEG-MI decoding and establishes a unified learning framework.

\textbf{Multi-channel spatial-spectral features} 
When performing different MI tasks, the amplitudes of EEG signals recorded by different electrodes can exhibit increases or decreases in specific spectral bands~\cite{altaheri2023deep}. The phenomenon is known as ERD and ERS. 
Therefore, it is critical to emphasize the relationship between different spectral features among multiple EEG electrodes.
However, the majority of existing work~\cite{schirrmeister2017deep, lawhern2018eegnet,miao2023lmda,li2021temporal,wimpff2024eeg,qin2024m} employs only a convolution with a kernel size of (channels, 1) to integrate information across multiple electrode channels, which is evidently insufficient for extracting effective spatial features. A smaller subset of studies has investigated the connectivity relationships among EEG channels using graph convolutional networks (GCNs) for EEG-MI.
Hou et al.~\cite{hou2022gcns} utilizes the absolute Pearson correlation matrix of EEG channels to quantify the correlation strength between electrodes, constructing a Laplacian graph for motor imagery EEG classification using GCNs. 
Meanwhile, SF-TGCN~\cite{tang2024spatial} constructs a learnable connectivity matrix for the GCN layer to model spatial features.
Nevertheless, few studies consider the unified modeling of the relationships among multiple spectral bands across various electrode channels.

In this work, we aim to comprehensively model the spatial-spectral relationships. We first extract multiple distinct spectral bands from each EEG channel to generate a multi-channel spatial-spectral representation. Then, we design a lightweight attention mechanism to reweight the channels based on their importance in the spatial-spectral dimension. This enables us to extract more efficient and discriminative spatial-spectral features. Compared to existing GCNs, our method inherently allows for finer-grained modeling of the spatial features present in EEG electrode channels.

\textbf{Long-term temporal features} 
 EEG signals contain rich temporal features, and substantial research efforts have also been devoted to extracting more effective temporal features from the EEG signals.
The transformer~\cite{vaswani2017attention}, with its inherent ability to model global information, has become a focal point in EEG-MI decoding. Recent studies~\cite{deny2023hierarchical,ahn2022multiscale,altaheri2022physics,song2022eeg,zhang2023local,liu2024msvtnet} have widely utilized transformers to extract long-term temporal features in MI-EEG, achieving good performance.
However, transformer models exhibit certain limitations, such as high parameters, high computational costs, and data-hungry characteristics, making them hard to use for real-time MI decoding.
In this work, we aim to develop a better method than the transformer for capturing long-term temporal features in EEG-MI decoding.

Inspired by recent advances~\cite{ding2022scaling, ding2024unireplknet} in computer vision, where large-kernel convolutions effectively capture long-term dependencies, we design a large-kernel variance pooling layer tailored for EEG signals. Specifically, the variance of EEG signals represents their spectral power~\cite{mane2020multi}, making it an effective operation for extracting temporal features. Based on the variance operation, we propose a multi-scale large-kernel variance pooling layer to capture long-term temporal features in EEG signals. This parameter-free and computationally efficient layer presents a promising alternative to transformers for extracting long-term temporal features in EEG decoding tasks.

\textbf{Limited data}
Some existing studies~\cite{perez2022eegsym,wei2024bdan,yin2024gitgan} employ transfer learning (TL) methods to enhance model generalization and alleviate small-sample challenges.
EEGsym~\cite{perez2022eegsym} uses extensive pretraining on other datasets followed by fine-tuning on the target dataset, while BDAN-SPD~\cite{wei2024bdan} combines data from other individuals with target data for co-training. Essentially, TL methods still rely on relatively large external data to improve generalization and involve complex training processes. Other studies adopt data augmentation to improve model performance with limited training samples. Conformer~\cite{song2022eeg} enhances data by segmenting and reconstructing signals in the time domain, though this may compromise signal continuity. Wang et al.~\cite{wang2024channel} augment MI data by simultaneously swapping channels and labels between the left and right brain regions, though this approach is limited to left and right MI tasks.

To address the above limitations, we propose a novel prototype learning (PL) approach from the perspectives of optimizing the feature space and the training process to enhance the model's generalization ability.
To the best of our knowledge, this is the first study to apply the PL to MI-EEG decoding. 
The classic PL method~\cite{yang2018robust} employs a prototype loss to push feature vectors towards corresponding prototypes, making the features within the same class more compact, which is beneficial for classification and model generalization. 
Based on the classic PL method, we develop a Dual Prototype Learning (DPL) approach to decouple inter-class separation and intra-class compactness in training processes. 
The DPL not only enhances intra-class compactness but also explicitly increases inter-class margins. 
More importantly, the DPL requires no additional training data and can serve as a general training strategy to enhance model generalization and help prevent overfitting.

In summary, from the perspectives of multi-channel spatial-spectral features, long-term temporal features, and the inherent small-sample dilemma of EEG-MI, we systematically identify the shortcomings of existing work. 
To overcome these limitations, we propose a novel end-to-end decoding framework for EEG-MI decoding, termed Spatial-Spectral and Temporal Dual Prototype Network (SST-DPN). The main contributions of this paper are as follows:
\begin{enumerate}
    \item We emphasize the importance of unified modeling of multi-channel spatial-spectral information in EEG-MI decoding, an aspect that has been overlooked in existing work. To this end, we design a lightweight attention mechanism that explicitly models the relationships among multiple channels in the spatial-spectral dimension. This method enables finer-grained spatial feature modeling, highlighting key spatial-spectral channels for the current MI task.
    \item To capture long-term temporal features from high temporal resolution EEG signals, we develop a Multi-scale Variance Pooling (MVP) module with large kernels. 
    The MVP has advantages over commonly used transformers for capturing long-term dependencies, as it is parameter-free and computationally efficient.
    \item To overcome the small-sample issue, we propose a novel DPL method to optimize feature space distribution, making same-class features more compact and different-class features more separated.
    The DPL acts as a regularization technique, enhancing the model's generalization ability and classification performance.
    \item We conduct experiments on three benchmark public datasets to evaluate the superiority of the proposed \method against state-of-the-art (SOTA) MI decoding methods. Additionally, comprehensive ablation experiments and visual analysis demonstrate the effectiveness and interpretability of our method.
    
\end{enumerate}

\section{Related Works}
\subsection{Deep Learning based EEG-MI Decoding}
With recent advancements in deep learning, researchers are increasingly using various deep learning architectures to decode EEG signals.
DeepConvNet~\cite{schirrmeister2017deep} employed multiple convolutional layers with temporal and spatial feature extraction kernels.
Sakhavi et al.~\cite{sakhavi2018learning} utilized FBCSP for feature extraction, followed by a convolutional neural network (CNN)-based classification.
Lawhern et al.~\cite{lawhern2018eegnet} introduced a compact network, EEGNet, employing depthwise and separable convolution for spatial-temporal feature extraction.
However, due to the lack of effective mechanisms for extracting highly discriminative features, the improvements of these methods are limited.

Attention mechanisms, which have recently gained significant recognition in various fields, have been successfully applied to MI-EEG decoding.
TS-SEFFNet~\cite{li2021temporal} combines a channel attention module based on the wavelet packet sub-band energy ratio with a temporal attention mechanism, followed by a feature fusion architecture.
LMDA-Net~\cite{miao2023lmda} combines a custom channel recalibration module with a feature channel attention module from ECA-Net~\cite{wang2020eca}.
Wimpff et al.~\cite{wimpff2024eeg} applied various attention mechanisms to the proposed BaseNet and made a very comprehensive comparison of the different variations.
M-FANet~\cite{qin2024m} uses a convolution with a small kernel size to extract local spatial information and a SE~\cite{hu2018squeeze} module to extract information from multiple perspectives.
These methods mainly apply attention mechanisms to deep features extracted by the neural network and improve the MI decoding accuracy to some extent.
Nonetheless, few studies have employed attention mechanisms to model the relationships between EEG electrode channels that reflect levels of brain activation~\cite{hanakawa2003functional}.

GCNs are considered promising for modeling relationships among EEG electrodes. While GCNs have been widely applied in EEG-based affective brain-computer interfaces, they remain relatively unexplored in motor imagery EEG signal recognition. Hou et al.~\cite{hou2022gcns} and Wang et al.~\cite{tang2024spatial} constructed graph networks using prior knowledge and learnable network parameters, respectively. These GCN approaches treat each EEG electrode as a node, with edges representing connections between electrodes. Although this approach captures spatial relationships across electrodes to some extent, it neglects the spectral features within each electrode, which are essential for effective EEG-MI recognition.

In comparison, extensive research efforts have focused on extracting more effective temporal features from high-temporal resolution EEG signals.
Lately, transformer models have made waves in natural language and computer vision due to the inherent perception of global dependencies~\cite{vaswani2017attention}. Transformers also emerged in MI decoding and achieved good performance, by leveraging long-term temporal relationships.
Conformer~\cite{song2022eeg} stacks transformer blocks to extract long-term dependency features based on local temporal features extracted by CNN.
ATCNet~\cite{altaheri2022physics} integrates the transformer with the temporal convolutional network (TCN) to extract long-range temporal features, resulting in good classification performance.
Similarly, MSVTNet~\cite{liu2024msvtnet} first employs the CNN to extract multi-scale local features and then uses the transformer to capture global features. Since the parameter count of the transformer module is significantly larger than that of the CNN module, MSVTNet introduces an auxiliary branch loss to prevent overfitting.
Nevertheless, an inherent conflict remains between the transformer model's need for large training datasets and the limited data available in EEG-MI tasks.
Additionally, transformer models have high parameters and computational costs, making them hard to use for real-time MI decoding.

\subsection{Prototype Learning}
PL simulates the way humans learn by memorizing typical examples to understand and generalize to new situations. 
In PL methods, a set of representative samples (prototypes) is learned during training, and during testing test samples are assigned to the closest prototype to determine their categories.
In~\cite{snell2017prototypical}, Snell et al. propose to apply the prototype concept for few-shot learning. 
However, this method learns the prototypes and the feature extractor separately using discriminative loss.
Yang et al.~\cite{yang2018robust} introduce the prototype model into the DL paradigm and design different discriminative loss as well as generative loss.
This significantly improves the performance and robustness of DL models in classification tasks.
Following the study~\cite{yang2018robust}, a substantial amount of research~\cite{yang2020convolutional,borlino2022contrastive,xia2023adversarial} has been devoted to using PL to learn a compact feature space for addressing open-world recognition.
Another line of research~\cite{zhang2021prototype,zhou2023revisiting} continues to explore the potential of PL in few-shot learning.
Prototype learning has demonstrated advantages in optimizing feature space to enhance model generalization, yet its application in EEG-MI decoding remains largely unexplored.

\section{Method}
\label{sec:method}
\begin{figure*}[t]
  \centering
  \vspace{-1cm}
  \includegraphics[width=1\textwidth]{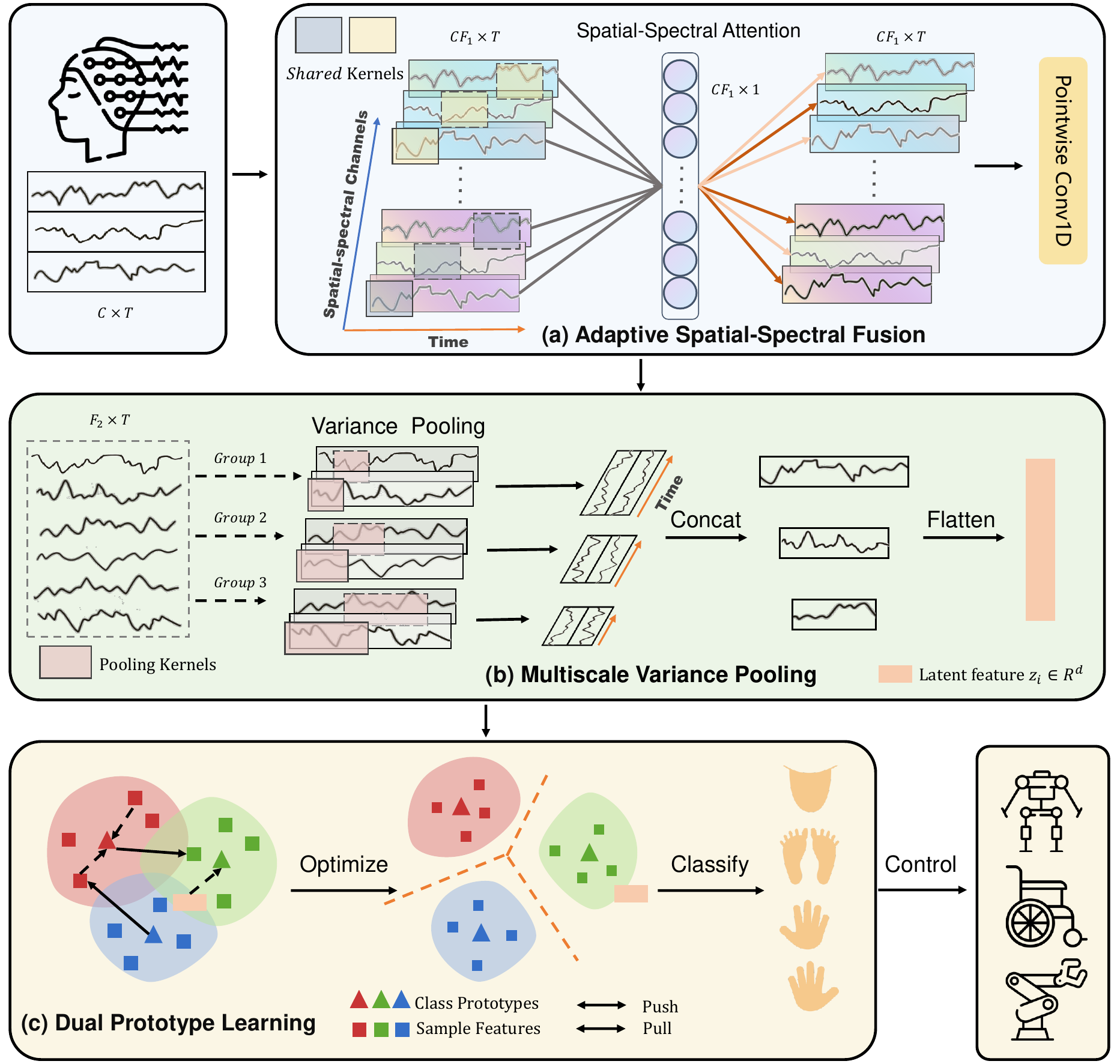}
  \caption{The overall framework of the proposed \method. Our \method takes raw EEG signals as input, extracts features through the ASSF and MVP modules, and makes classification decisions via the DPL module, enabling its application in human-computer interaction control.}
\vspace{-0.5cm}
  \label{framework}
\end{figure*}

Our proposed \method is an end-to-end 1D convolutional architecture consisting of three modules, as shown in \figurename~\ref{framework}.
The Adaptive Spatial-Spectral Fusion (ASSF) module first employs LightConv to extract multiple spectral bands from each EEG electrode, then utilizes spatial-spectral attention (SSA) to highlight important spatial-spectral channels, and finally applies a pointwise convolution to fuse features across channels.
The multi-scale variance pooling module leverages large-kernel variance pooling at different scales to extract long-term temporal features.
The dual prototype learning module is used to optimize the feature space and make classification decisions.
In this section, we will provide a detailed introduction to each module.

\subsection{EEG Representation}
In this paper, we feed raw MI-EEG signals into the proposed model without any additional preprocessing.
Given a set of $m$ labeled MI trials $ S =\{X_{i},y_{i}\}_{i=1}^{m} $, where $ X_i\in\mathbb{R}^{C\times T}$ consists of $ C $ channels (EEG electrodes) and $T$ time points, $ y_i \in \{1,...,n \}$ is the corresponding class label, and $ n $ is the total number of predefined classes for set $ S $, our \method model first maps a motor imagery trail $X_i$ to the feature space $ Z $ and obtains $ z_i = f(X_i)\in\mathbb{R}^{d}$, where $ f $ is the feature extractor composed of ASSF and MVP, as shown in \figurename~\ref{framework}. 
Then, the DPL module maps the feature $ z_i $ to its corresponding class $ y_i $.

\subsection{Adaptive Spatial-Spectral Fusion}
\subsubsection{LightConv for Spatial-Spectral Representation}
Since different MI classes may differ in their corresponding spectral-spatial sensorimotor rhythm (SMR) patterns~\cite{lawhern2018eegnet}, most existing studies first extract multi-view spectral information from each EEG electrode channel to form a spatial-spectral representation.
In this paper, we use a 1D convolution, LightConv, to extract spatial-spectral features from raw EEG signals.
The LightConv first divides the input signal $X_i\in\mathbb{R}^{C\times T}$ into $h$ groups along the channel dimension. 
Therefore, each group has $C/h$ channels, and each channel within the same group shares convolutional weights.
The implementation steps are as follows:
\begin{align}
    X_h &= \textrm{Reshape}(X_i) \in \mathbb{R}^{{(C/h)} \times h \times T}\label{eq1}\\
    X_{dw} &= \textrm{DWConv1D}(X_h,  W)  \in \mathbb{R}^{{(C/h)} \times {(h*F_1)} \times T}\label{eq2}\\
    X_{ssr} &= \textrm{Reshape}(X_{dw}) \in \mathbb{R}^{{CF_1} \times T}\label{eq3},
\end{align}
where $ \textrm{DWConv1D} $ denotes 1D depthwise convolution. Additionally, $ W\in\mathbb{R}^{(h*F_1) \times 1 \times k} $ is the learnable convolution parameter, and $X_{ssr}\in\mathbb{R}^{CF_1\times T} $ corresponds to the resultant spatial-spectral representation. 
It is important to note that $ W $ contains $ h*F_1 $ filters, as each channel uses $ F_1 $ filters with kernel size $k$ to generate different spectral characteristics.
In this paper, we set $ h $ as 1. All electrode channels share $ F_1 $ temporal filters to produce spatial-spectral representation $X_{ssr}$, as shown in \figurename~\ref{framework}.

\subsubsection{Spatial-Spectral Attention}
After obtaining the spatial-spectral representation, we can model the relationships between EEG electrode channels at a finer granularity. Considering that different MI classes exhibit distinct spectral-spatial activation patterns, each channel should not be treated equally; instead, emphasis should be placed on regions and spectral bands relevant to the specific MI task.
Attention mechanisms are well-suited for this purpose. Thus, we designed the SSA to enhance the extraction of powerful spatial-spectral features.

\begin{figure}[tbp]
  \centering
    \vspace{-1cm}
  \includegraphics[width=0.95\columnwidth]{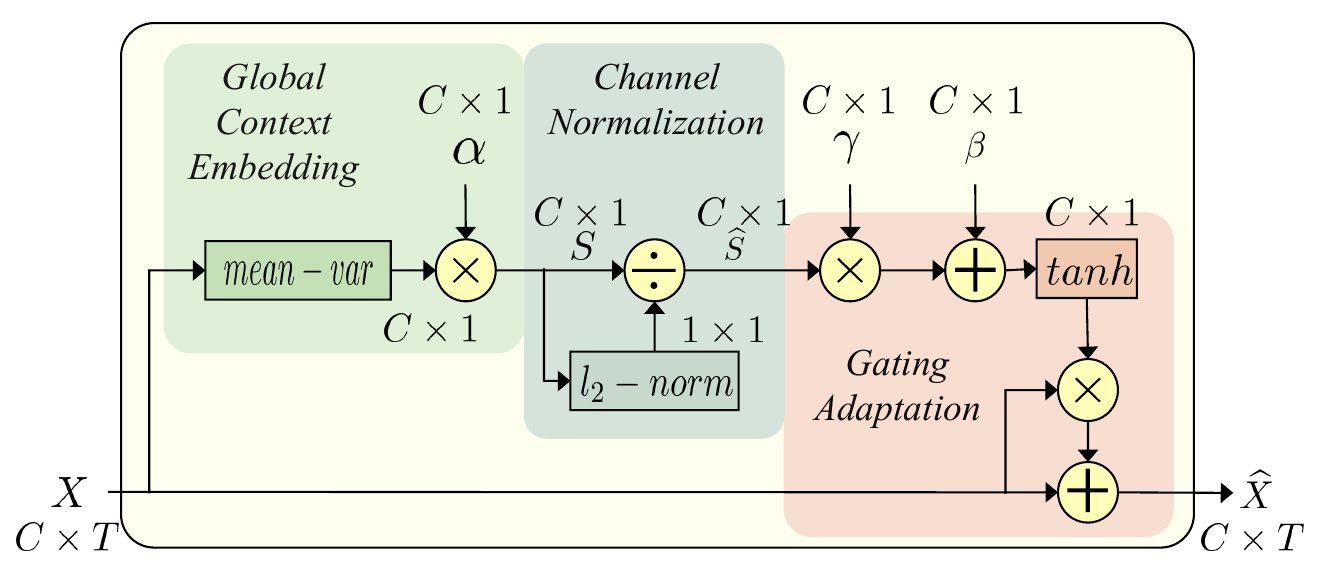}
  \caption{The detailed diagram of the implementation of spatial-spectral attention.}
  \label{fig2}
\end{figure}

Inspired by the gated channel transformation~\cite{yang2020gated}, our spatial-spectral attention consists of three parts: global context embedding, channel normalization, and gating adaptation, as shown in \figurename~\ref{fig2}. For an input $X\in\mathbb{R}^{C\times T}$, global context embedding employs a $\textrm{mean-var}$ operation to aggregate temporal information from each channel. This involves calculating the variance within each 1-second window, followed by averaging them. Subsequently, $ \alpha$ is responsible for controlling the weight of each channel:
\begin{equation}
    s = \alpha \cdot \textrm{mean-var}(X).
\end{equation}
Then, we use a channel normalization component to model channel relations:
\begin{equation}
\hat{s}=\frac{\sqrt{C}s}{||\mathbf{s}||_{2}}=\frac{\sqrt{C}s}{[(\sum\limits_{c=1}^{C}s^{2})+\epsilon]^{\frac{1}{2}}},
\end{equation}
where $\epsilon$ is a small constant to avoid the problem of derivation at the zero point. The gating weight and bias, $\gamma$ and $\beta$ are responsible for adjusting the scale of the input feature channel-wise:
\begin{align}
&\textrm{Attention} = 1+\tanh(\gamma\hat{s}+\beta)\label{eq6}\\
&\hat{X}=X \cdot \textrm{Attention}.
\end{align}

The scale of each channel of $X_{ssr}\in\mathbb{R}^{CF_1\times T} $ output by the SSE module will be adjusted by the corresponding attention weight. Additionally, SSA leverages global temporal information to model channel relationships and modulate feature maps on the channel-wise level. Therefore, we can effectively fuse the weighted spatial-spectral features using a simple 1D pointwise convolution. As shown in \figurename~\ref{framework}, the pointwise convolution uses $ F_2 $ filters to simultaneously fuse spectral features of all electrodes to get $X_{assf}\in\mathbb{R}^{F_2\times T} $.

\subsection{Multi-scale Variance Pooling}
It is crucial to acquire long-term dependencies and global temporal information for MI-EEG decoding. 
Transformer models can capture global information well by using the self-attention mechanism, but they have a large number of parameters and high computational complexity. 
In fact, under the constrained EEG training data, it is difficult for the transformer-based model to achieve optimal performance as in the computer vision field. 
Therefore, it is necessary to design a new method for EEG decoding that can extract long-term temporal features.

Recent work~\cite{ding2022scaling, ding2024unireplknet} in computer vision has demonstrated that large-kernel convolutions have the potential to capture long-term dependency.
Additionally, Metaformer~\cite{yu2023metaformer} has proposed that using a simple pooling layer in place of self-attention in transformers can also perform well. 
Combining these two points, we consider using a pooling layer with a large kernel to extract the long-term temporal information from the EEG signals. 
Furthermore, previous work~\cite{mane2020multi} indicated that the variance of EEG signals represents their spectral power, which can serve as an important feature for distinguishing different MI classes.
Building on these insights, we develop the multi-scale variance pooling module tailored to effectively capture long-range temporal features for MI-EEG signals.

\begin{figure}[!tb]
  \centering
    \vspace{-1cm}
  \includegraphics[width=0.95\columnwidth]{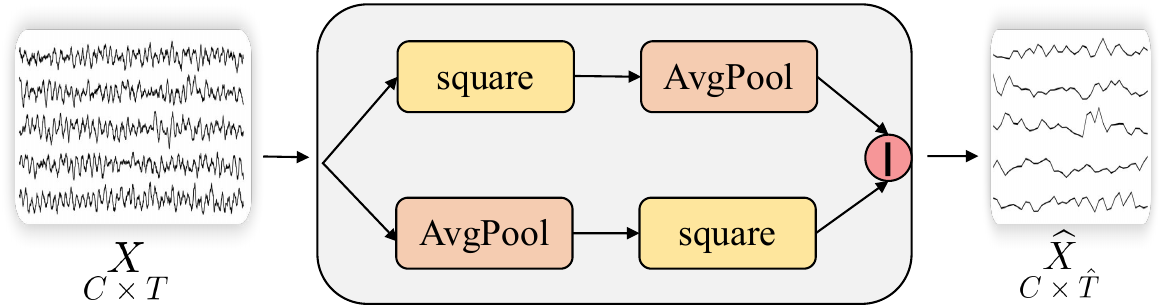}
  \caption{An illustration of the proposed VarPool layer. }
  \label{fig3}
\end{figure}
First, we design a 1D variance pooling layer, denoted as $\textrm{VarPool}$.
For a time series signal $x\in\mathbb{R}^{t}$, the relationship between its variance $Dx$ and mean $Ex$ is as follows:
\begin{align}
    Dx &= E(x-Ex)^2\nonumber\\
    &=E(x^2)+(Ex)^2-2*(Ex)^2\nonumber\\
    &=E(x^2)-(Ex)^2.
\end{align}
Therefore, for the EEG representation $X\in\mathbb{R}^{C\times T} $, we can utilize average pooling to calculate variance pooling, as shown in \figurename~\ref{fig3}:
\begin{equation}
    \textrm{VarPool}(X)_{k,s} = \textrm{AvgPool}(X^2)_{k,s} - \textrm{AvgPool}(X)_{k,s}^2 ,
\end{equation}
where $k$ and $s$ represent the specified window length and sliding step size. For an input $X\in\mathbb{R}^{C\times T} $, the VarPool layer slides along the time dimension to calculate the variance within each window to obtain the output $\hat{X}\in\mathbb{R}^{C\times \hat{T}} $ :
\begin{equation}
    \hat{T} = \left\lfloor\frac{T+2\times\text{padding}-(k-1)-1}{s}+1\right\rfloor .
\end{equation}

Furthermore, we incorporate a multi-scale design into the variance pooling layer, forming our MVP module.
Specifically, the output $X_{assf}$ of the ASSF module is split into three groups along the channel dimension. VarPool layers with different large kernel sizes (i.e., 50, 100, and 200) are used for each group to extract temporal features. 
Then, the outputs of the three groups are flattened and concatenated to obtain the final feature vector $ z_i $. 
It is noteworthy that our MVP module contains no trainable parameters and is computationally efficient.

\begin{figure}[thbp]
\vspace{-1cm}
\centering
\subfigure[CE loss]{
\includegraphics[width=2.50in]{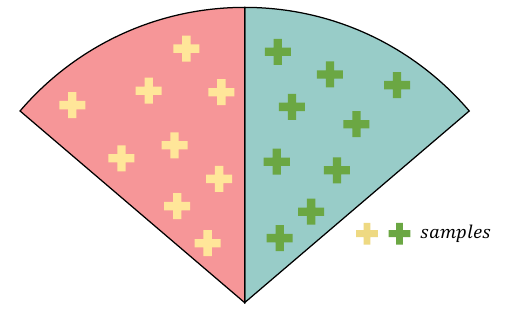}
}
\subfigure[PL]{
\includegraphics[width=2.50in]{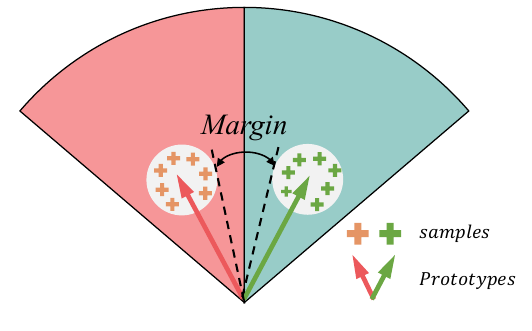}
}
\subfigure[One way to improve PL]{
\includegraphics[width=2.50in]{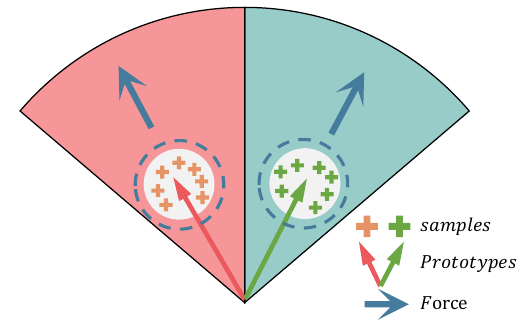}
}
\subfigure[DPL]{
\includegraphics[width=2.50in]{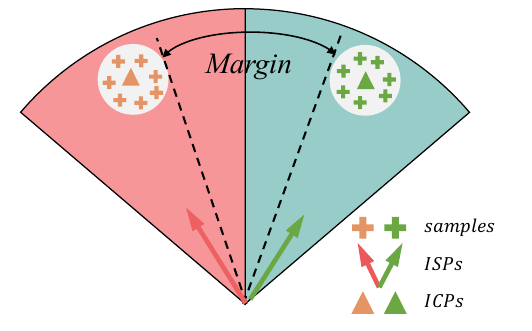}
}
\caption{An illustration of the feature space distribution for CE loss, PL, and our proposed DPL. CE loss only ensures that features of different classes are generally separable, resulting in a loose feature space. PL increases the compactness of features within the same class. Building on PL, our DPL further enlarges the margins between features of different classes. }
\label{fig4}
\end{figure}
\subsection{Dual Prototype Learning}
In classification tasks, existing methods input the feature vector $z$ into a multi-layer perceptron (MLP) to obtain classification results, and optimize the parameters of the neural network using cross-entropy (CE) loss. Recent studies~\cite{ahn2022multiscale, liu2022fbmsnet} indicate that CE loss may be lacking in the effectiveness of reducing intra-class variation, especially when considering the non-stationarity of EEG signals. 
As shown in \figurename~\ref{fig4} (a), CE loss only optimizes samples towards the decision boundary of the corresponding class, resulting in a loose distribution of sample features.
When applying PL to classification, the first step is to assign a prototype to each class.
A classification loss optimizes the sample features to be closest to its corresponding prototypes for classification. Additionally, a prototype loss is used to further push the sample features towards its corresponding prototypes, which can increase intra-class compactness as shown in \figurename~\ref{fig4} (b), while also acting as a form of regularization to prevent model overfitting.

Although PL has been widely applied in the field of computer vision, its potential in EEG-MI decoding has been scarcely researched. 
PL methods focus on utilizing prototype loss to increase intra-class compactness, thereby implicitly enhancing inter-class distance to form a margin, as illustrated in \figurename~\ref{fig4} (b). 
Benefitting from larger margins, the PL methods outperform CE loss in both general classification tasks~\cite{yang2018robust,yang2020convolutional} and few-shot learning~\cite{snell2017prototypical}.
Therefore, in this paper, we are dedicated to further increasing inter-class margins based on the PL method, in order to enhance 
the model's generalization capability in MI decoding tasks with small samples.
A natural idea is to extend the clustered features along the direction of their corresponding prototypes, as shown in \figurename~\ref{fig4} (c). 
We achieve this goal using Dual Prototype Learning, ultimately obtaining a larger inter-class margin as shown in \figurename~\ref{fig4} (d).

Specifically, we develop two prototypes for each class: the Inter-class Separation Prototype (ISP) and the Intra-class Compact Prototype (ICP), aiming to achieve inter-class separation and intra-class compactness, respectively.
Both ISPs and ICPs are learnable vectors with the same shape as \( z_i \) in \figurename~\ref{framework} and are randomly initialized at the start of training.
Based on the ISPs, we utilize softmax and CE loss to achieve inter-class separation:
\begin{equation}
\mathcal{L}_{S}(s,z)=-\frac{1}{m}\sum_{i=1}^{m}log\frac{e^{s_{y_{i}} \cdot {z_i}}}{\sum_{j=1}^{n}e^{s_j \cdot z_i}},
\label{eq11}
\end{equation}
where $m$ is the number of training samples, $n$ is the number of classes, $z_i$ is the feature of the $i$-th sample, $y_i$ is the corresponding label in range $[1,n]$, $s$ represents the ISPs, and $s_j\in\mathbb{R}^{d}$ is the ISP of class $j$. 
Minimizing $\mathcal{L}_S$ can facilitate the separation of features from different classes, resulting in a feature space similar to \figurename~\ref{fig4} (a).

Furthermore, we use intra-class compactness loss to compress the distance between samples belonging to the same class in the feature space, which is defined as:
\begin{equation}
    \mathcal{L}_C(c,z)=\sum_{i=1}^m{D(z_i, c_{y_i})},
    \label{eq12}
\end{equation}
where $c$ represents the ICPs, $c_{y_i}\in\mathbb{R}^{d}$ is the ICP of class $y_i$, and $D$ is the distance function.
To prevent training oscillations and mitigate the influence of outlier samples, we use the Huber loss $\mathcal{L}_\delta(z,c)$ with $\delta=1$ as the distance function $D(z, c)$, which is defined as:
\begin{equation}
\mathcal{L}_\delta(z,c)=\begin{cases}\frac{1}{2}(z-c)^2&\text{if}\quad|z-c|\leq\delta\\
\delta|z-c|-\frac{1}{2}\delta^2&\text{if}\quad|z-c|> \delta\end{cases}.
\label{eq13}
\end{equation}
$\mathcal{L}_C $ can represent the compactness of each class's feature vectors. 
By minimizing $\mathcal{L}_C $, we can increase the intra-class compactness so that features of the same class are clustered together like \figurename~\ref{fig4} (b).

Previous PL methods use a single prototype for each class, as described in \figurename~\ref{fig4} (b). In comparison, we decouple inter-class separation and intra-class compactness by using ISPs and ICPs.
On one hand, this decouple enhances the robustness of the training process. On the other hand, it provides the conditions for further increasing the inter-class margins.
Specifically, we apply an implicit force and an explicit force to the features to achieve the feature space optimization from \figurename~\ref{fig4} (c) to \figurename~\ref{fig4} (d).
\begin{itemize}
    \item \textbf{Implicit force.} Due to the softmax's properties~\cite{wang2017normface}, the $\mathcal{L}_S $ tends to increase $s_{y_{i}} \cdot {z_i}$ until it converges to a constant value during training. 
    This procedure simultaneously displaces the feature vector $z_i$ and its corresponding ISP $s_{y_i}$ away from the origin of the feature space until convergence. 
    Furthermore, if we constrain the norm of ISPs to a smaller value, i.e., $\left\|s_i\right\|_2\leq S$ (weight-normalization), the feature vectors will be pushed further away from the origin. 
    This constraint can act as an implicit force.
    \item \textbf{Explicit force.} To complement the implicit force, we design a simple loss function, $\mathcal{L}_{EF}$,  to increase the norms of ICPs:
    \begin{equation}
    \mathcal{L}_{EF}(c)=-\Vert c \Vert_2.
    \label{eq14}
    \end{equation}
    By minimizing $\mathcal{L}_{EF}$, the norms of ICPs increase, thereby guiding the features to be pushed away from the origin.
\end{itemize}

During the training phase, the optimization objective of the proposed DPL is as follows:
\begin{equation}
\renewcommand{\arraystretch}{1.5}
\begin{array}{ll}
\operatorname{minimize} & \mathcal{L}_S(s,z) + {\lambda}_1\mathcal{L}_C(c,z) + {\lambda}_2\mathcal{L}_{EF}(c) \\
\text {subject to } & \left\|s_i\right\|_2\leq S, \quad \forall i=1,2, \ldots n
\end{array},
\label{eq15}
\end{equation}
where ${\lambda}_1$ and ${\lambda}_2$ are the trade-off scalar to balance the three losses, and $S$ is set to 1.
During the testing phase, the test sample $X_i$ is classified by calculating the dot product between its feature vector $z_i$ and the ISP for each class:
\begin{equation}
\hat{y}_i=\underset{j}{\operatorname*{argmax}}(z_i\cdot s_j),\quad  j=1,2, \ldots n,
\label{eq16}
\end{equation}
where $\hat{y}_i$ denotes the predicted result.

\section{Experiments and results}
\label{sec:Experiments}
\subsection{Evaluation Datasets}
To demonstrate the effectiveness of our \method, we evaluate it on two public MI-EEG datasets, namely, BCI4-2A~\cite{tangermann2012review} and BCI4-2B~\cite{ang2012filter}. Additionally, we use the BCI3-4A~\cite{blankertz2006bci} with fewer training data to further validate the generalization ability of the proposed method. The detailed information of the three datasets is presented in Table~\ref{dataset}.

\begin{table*}[htbp]
\centering
\caption{Summary of the public datasets used in the experiments. We denote BCI4-2A, BCI4-2B, and BCI3-4A as Dataset I, Dataset II, and Dataset III, respectively.}
\label{dataset}
\renewcommand{\arraystretch}{1.20}
\begin{tabular}{@{}lccccccc@{}}
\toprule
Datasets    & EEG electrodes & Subjects & Trials/subject & classes & Sampling rate (Hz) & Duration (s) & Data split \\ \midrule
Dataset I   & 22             & 9        & 576                & 4       & 250               & 4            & Official   \\
Dataset II  & 3              & 9        & 720                & 2       & 250               & 4            & Official   \\
Dataset III & 118            & 5        & 280                & 2       & 100               & 3.5          & Official   \\ \bottomrule
\end{tabular}

\end{table*}

\textbf{Dataset I.} BCI4-2A provided by Graz University of Technology consists of EEG data from 9 subjects. There were four motor imagery tasks, covering the imagination of moving the left hand, right hand, both feet, and tongue. Two sessions on different days were collected with 22 Ag/AgCl electrodes at a sampling rate of 250 Hz. One session contained 288 EEG trials, i.e., 72 trials per task. We use [2, 6] seconds of each trial and all 22 electrodes in the experiments.

\textbf{Dataset II.} BCI4-2B provided by Graz University of Technology consists of EEG data from 9 subjects. There were two motor imagery tasks, covering the imagination of moving left and right hand. Five sessions were collected with three bipolar electrodes (C3, Cz, and C4) at a sampling rate of 250 Hz and each session contained 120 trials. We use the [3, 7] seconds of each trial in the experiments.

\textbf{Dataset III.} BCI3-4A, recorded at 100 Hz using 118 electrodes, contains 280 trials per subject and comprises two distinct classes: right hand, and foot. This dataset distinguishes itself from other datasets through its imbalanced division into training and testing trials. The quantity of training trials fluctuates between 28 and 224, varying with the subject (al: 224, aa: 168, av: 84, aw: 56, ay: 28), with the residual trials designated for testing. Each trial lasts 3.5 seconds. To preclude overfitting by reducing the number of data points per trial, we select the three channels shared (C3, Cz, and C4).

As the competition guidelines~\cite{ang2012filter} for BCI4-2A and BCI4-2B datasets, we apply hold-out analysis to evaluate the performance of our \method and all comparison methods. As such, the model is trained and tested completely in different sessions. This evaluation method is more in line with practical application scenarios and can better test the generalization ability of the model.
For the BCI3-4A dataset, we follow its official protocol to further validate the advantages of our method on small-sample training datasets.
\begin{table*}[!t]
\centering
\caption{Classification accuracy(\%) and kappa comparisons with SOTA methods on Dataset I. }
\label{dataset1}
\resizebox{\textwidth}{!}{
\begin{threeparttable}
\renewcommand{\arraystretch}{1.18} 
\begin{tabular}{@{}lccccccccccccc@{}}
\toprule
Methods    & A01   & A02   & A03   & A04   & A05   & A06   & A07   & A08   & A09   & Average & Std   & Kappa  & p-value              \\ \midrule
FBCSP~\cite{ang2008filter}      & 76.00 & 56.50 & 81.25 & 61.00 & 55.00 & 45.52 & 82.75 & 81.25 & 70.75 & 67.75   & 13.67 & 0.5700 & 0.0020               \\
EEGNet~\cite{lawhern2018eegnet}     & 85.76 & 61.46 & 88.64 & 67.01 & 55.90 & 52.08 & 89.58 & 83.33 & 79.51 & 74.50   & 14.69 & 0.6600 & 0.0020               \\
TS-SEFFNet~\cite{li2021temporal} & 82.29 & 49.79 & 87.57 & 71.74 & 70.83 & 63.75 & 82.92 & 81.53 & 81.94 & 75.17   & 12.01 & 0.6630 & 0.0020               \\
LMDA-Net~\cite{miao2023lmda}   & 83.90 & 60.30 & 88.10 & \underline{78.20} & 56.20 & 57.20 & 88.40 & 82.70 & 84.30 & 75.40   & 13.56 & 0.6700 & 0.0020               \\
Basenet-SE~\cite{wimpff2024eeg} & 81.60 & 52.08 & 90.28 & 73.96 & 76.39 & 62.85 & 86.81 & 80.56 & 79.51 & 76.00   & 11.91 & 0.6794 & 0.0020               \\
M-FANet~\cite{qin2024m}    & 86.81 & \textbf{75.00} & 91.67 & 73.61 & 76.39 & 61.46 & 85.76 & 75.69 & 87.17 & 79.28   & 9.94  & 0.7259 & 0.0137               \\
Conformer~\cite{song2022eeg}  & 87.85 & 54.86 & 86.46 & 76.04 & 58.33 & 59.72 & 89.58 & 83.33 & 81.25 & 75.27   & 13.85 & 0.6702 & 0.0020               \\
ATCNet~\cite{altaheri2022physics}     & 86.81 & 68.40 & 92.01 & 73.61 & \underline{78.82} & 62.15 & 86.46 & 87.15 & 83.33 & 79.86   & 9.94  & 0.7312 & 0.0020               \\
MSVTNet~\cite{liu2024msvtnet}     & 80.56 & 54.51  & 92.36 & 73.96 & 75.69 & 63.89 & 76.39 & 82.99 & 79.17 & 75.50   & 10.94  & 0.6733 & 0.0020  \\
FBMSNet~\cite{liu2022fbmsnet}    & 87.85 & 66.32 & 92.36 & 76.74 & 72.57 & 62.15 & 80.21 & 86.46 & \underline{87.85} & 79.17   & 10.51  & 0.7235 & 0.0020               \\
BDAN-SPD~\cite{wei2024bdan}   & \underline{88.97} & 57.24 & 92.67 & 74.54 & 55.80 & 58.60 & \textbf{93.14} & \underline{88.93} & 87.50 & 77.49   & 15.22 & 0.6998 & 0.0195               \\
SF-TGCN~\cite{tang2024spatial}   & 84.44 & 68.69 & \textbf{93.35} & 77.94 & 77.50 & \textbf{71.22} & 87.39 & 84.48 & 82.21 & \underline{80.82}   & \textbf{7.79} & \underline{0.7440} & 0.0098               \\
\textbf{\method}     & \textbf{89.58} & \underline{71.88} & \underline{93.06} & \textbf{82.64} & \textbf{81.25} & \underline{70.14} & \underline{89.93} & \textbf{89.24} & \textbf{89.24} & \textbf{84.11}   & \underline{8.30}  & \textbf{0.7881} & - \\ \bottomrule
\end{tabular}
\begin{tablenotes} 
\item Best performances are highlighted in bold, while the second-best with underlined.
\end{tablenotes} 
\end{threeparttable}  
}
\end{table*}

\subsection{Experimental Setups}
 \subsubsection{Experimental Details}
In this study, we implement our \method using the PyTorch library, based on Python 3.10 with an Nvidia Geforce 2080Ti GPU. 
We use the AdamW optimizer with default settings ($\textrm{learning rate} = 0.001, \textrm{weight decay}=0.01$) to train the feature extractor of our \method. 
Additionally, we use another Adam optimizer to optimize ISPs and ICPs, with a learning rate of 0.001 on Datasets I and II, and a learning rate of 0.01 on the small-sample Dataset III.
Moreover, for the hyperparameters of the model in \figurename~\ref{framework}, on Dataset I and II, we empirically set the kernel size of LightConv as 75, $F_1$ and $F_2$ as 9 and 48, and the kernel sizes of different scales of the MVP layer as 50, 100, and 200.
On Dataset III, due to the differences in sampling rate, we set the kernel size of LightConv as 50, and the kernel sizes of different scales of the MVP layer as 50, 100, and 150.

To prevent overfitting and reduce the number of epochs needed to train the model, a two-stage training strategy as in~\cite{schirrmeister2017deep} is used in this work. Specifically, during the training phase, the training data is split into a training set and a validation set. In the first stage, only the training set is used, and the training is stopped if there is no decrease in the validation set loss for $N_e$ consecutive epochs or reach the maximum training epoch $N_1$. During the second training stage, all training data are employed, then continue training for $N_2$ epochs.
We use the model from the final epoch of the training phase to obtain evaluation metrics on the test set.
Due to the different sizes of the datasets used, we set $N_1$, $N_e$, and $N_2$ to be 1000, 200, and 300 respectively for Dataset I, and 300, 150, and 200, respectively, for Dataset II. For Dataset III, we set them to be 300, 150, and 150.

 \begin{table*}[th]
\centering
\caption{Classification accuracy(\%) and kappa comparisons with SOTA methods on Dataset II.}
\label{dataset2}
\resizebox{\textwidth}{!}{
\begin{threeparttable}
\renewcommand{\arraystretch}{1.18} 
\begin{tabular}{@{}lccccccccccccc@{}}
\toprule
Methods    & B01   & B02   & B03   & B04   & B05   & B06   & B07   & B08   & B09   & Average & Std   & Kappa  & p-value \\ \midrule
FBCSP~\cite{ang2008filter}      & 70.00 & 60.36 & 60.94 & \underline{97.50} & 93.12 & 80.63 & 78.13 & 92.50 & 86.88 & 80.01   & 13.85 & 0.6000 & 0.0059  \\
EEGNet~\cite{lawhern2018eegnet}     & 71.50 & 58.65 & 81.12 & 96.25 & 86.23 & 77.88 & 85.12 & 91.10 & 80.15 & 80.89   & 11.06 & 0.6321 & 0.0020  \\
TS-SEFFNet~\cite{li2021temporal} & 72.81 & 65.71 & 75.75 & 96.25 & 91.25 & 85.00 & 88.63 & 91.87 & 82.18 & 83.27   & 10.08  & 0.6637 & 0.0020  \\
LMDA-Net~\cite{miao2023lmda}   & 75.80 & 63.20 & 65.20 & 97.30 & 94.30 & 84.50 & 82.40 & 92.90 & 87.00 & 82.51   & 12.28 & 0.6500 & 0.0039  \\
Basenet-SE~\cite{wimpff2024eeg} & 72.50 & \underline{67.86} & \underline{81.13} & 96.86 & 93.44 & 84.69 & \underline{88.75} & \underline{93.44} & 84.69 & 84.82   & 9.76  & 0.6918 & 0.0057  \\
M-FANet~\cite{qin2024m}    & 73.75 & 66.07 & \textbf{82.81} & 94.69 & 89.69 & 84.06 & 88.12 & 90.62 & 88.75 & 84.28   & \textbf{9.06}  & 0.6857 & 0.0125  \\
Conformer~\cite{song2022eeg}  & 74.56 & 57.00 & 62.50 & 97.01 & 92.36 & 83.44 & 85.00 & \underline{93.44} & 87.19 & 81.39   & 13.96 & 0.6265 & 0.0086  \\
ATCNet~\cite{altaheri2022physics}     & 72.50 & 67.64 & 80.31 & 95.94 & \underline{96.06} & \underline{88.12} & 86.88 & 89.69 & \underline{90.94} & \underline{85.34}   & 9.94  & \underline{0.7068} & 0.0645  \\
MSVTNet~\cite{liu2024msvtnet}     & 71.56 & 66.43  & 80.63 & 96.25 & \textbf{96.25} & 86.25 & \underline{88.75} & 90.31 & 89.38 & 85.09   & 10.37  & 0.7018 & 0.0250  \\
FBMSNet~\cite{liu2022fbmsnet}    & 71.30 & 55.20 & 80.55 & 97.15 & 95.00 & 84.66 & 85.23 & 91.10 & 87.13 & 83.04   & 12.99 & 0.6692 & 0.0039  \\
BDAN-SPD~\cite{wei2024bdan}   & \textbf{83.02} & 63.27 & 62.87 & \textbf{98.37} & 89.74 & \textbf{92.84} & 87.07 & \textbf{96.09} & \textbf{93.47} & 85.19   & 13.36 & 0.6916 & 0.4101  \\
\textbf{\method}     & \underline{77.50} & \textbf{68.93} & \textbf{82.81} & 96.88 & \textbf{96.25} & 87.50 & \textbf{89.06} & \underline{93.44} & 87.50 & \textbf{86.65}   & \underline{9.10}  & \textbf{0.7330} & -       \\ \bottomrule
\end{tabular}

\begin{tablenotes} 
\item Best performances are highlighted in bold, while the second-best with underlined.
\end{tablenotes} 
\end{threeparttable}}
\end{table*}
\subsubsection{Comparison methods}
We compare our \method with twelve methods\footnote{Most existing studies have only been tested on Datasets I and II, while the SF-TGCN paper only presents results for Dataset I. For Dataset III, we made efforts to reproduce applicable methods and present the results.}, including two baseline methods, FBCSP~\cite{ang2008filter} and EEGNet~\cite{lawhern2018eegnet}, as well as ten competitive SOTA methods from the past three years.
Among these methods, TS-SEFNet~\cite{li2021temporal}, LMDA-Net~\cite{miao2023lmda}, Basenet-SE~\cite{wimpff2024eeg}, and M-FANet~\cite{qin2024m} are attention-based approaches that fully explore the application of attention mechanisms in the deep feature dimension. Conformer~\cite{song2022eeg}, ATCNet~\cite{altaheri2022physics}, and MSVTNet~\cite{liu2024msvtnet} leverage transformers to capture long-range temporal features in EEG signals. FBMSNet~\cite{liu2022fbmsnet} employs mixed convolutions to extract multi-scale features and incorporates center loss to enhance intra-class compactness. BDAN-SPD~\cite{wei2024bdan} uses a transfer learning approach, utilizing large amounts of data from other subjects to aid in training.
SF-TGCN~\cite{tang2024spatial} leverages GCN to learn the topological connections between electrodes.
\subsubsection{Performance Metrics}
In the experiments, the classification accuracy (ACC) and Cohen's kappa coefficient (Kappa) are used as two metrics for performance evaluation. The mathematical formula of Cohen's kappa coefficient is defined as follows:
\begin{align}
Kappa & =\frac{P_0-P_e}{1-P_e},
\end{align}
where $p_0$ represents the classification accuracy of the model and $p_e$ represents the expected consistency level.
Nevertheless, a one-sided Wilcoxon signed-rank test is used to verify the significance of improvement. 

\subsection{Overall Performance Comparison}

We conduct extensive experiments and compare our method with numerous SOTA approaches across three public datasets. 
\tablename~\ref{dataset1} displays the classification performance of all methods on Dataset I. Our \method method achieves the highest average accuracy of 84.11\% and the highest average kappa value of 0.7881. Moreover, Our method achieves a level of significance with $\textrm{p} < 0.05$ compared to all benchmark methods.
The results demonstrate that the classification accuracy of our proposed \method is not only 16.36\% higher than the competition champion solution FBCSP ($\textrm{p} < 0.01$) but also significantly surpasses the classic EEGNet by 9.61\% ($\textrm{p} < 0.01$). 
The four latest attention-based methods all apply the attention mechanism to deep feature dimensions. In contrast, our \method utilizes lightweight attention in the spatial-spectral dimension. Consequently, our approach is more interpretable and has an accuracy of 4.83\% higher than the best attention-based method M-FANet. 
Compared to transformer-based methods, we utilize a non-parametric and computationally efficient MVP module to extract long-term temporal features. The results demonstrate that our method achieves accuracy improvements of 4.25\%, 8.84\%, and 8.61\% compared to ATCNet, Conformer, and MSVTNet, respectively.
Compared to FBMSNet, our \method not only increases intra-class compactness but also further enlarges inter-class margins, displaying an accuracy improvement of 4.94\%.
Compared to SF-TGCN, our \method models spatial features at a finer granularity and comprehensively addresses challenges in EEG-MI decoding, resulting in a 3.29\% improvement in accuracy.
In contrast to BDAN-SPD, our \method does not require a large amount of data from other subjects for training, and it achieves a 6.62\% higher accuracy.

The experimental results on Dataset II are presented in \tablename~\ref{dataset2}, which shows similar results to those on Dataset I. Our \method also achieves the highest average accuracy of 86.65\% with the smallest standard deviation and the highest average kappa value of 0.7330. Our method demonstrates significant advantages compared to most of the comparison methods.
Notably, on Datasets I and II, our method achieves the best or second-best accuracy for nearly all subjects. Particularly, on Dataset I, for subjects A02 and A06 where other methods do not perform well, our method achieves an accuracy above 70\%.
As suggested in~\cite{liu2022fbmsnet}, a BCI system with $> 70\%$ binary classification accuracy is generally considered to be usable for healthy subjects and stroke patients.
This demonstrates the potential of our \method for MI-based BCI applications.

Moreover, on the smaller training Dataset III, we reproduce several SOTA methods suitable for comparison on this dataset. 
As shown in \tablename~\ref{dataset3}, the number of training samples for the 5 subjects in Dataset III decreases from 224 to 28.
Our \method achieves an average recognition accuracy of 82.03\% and a kappa value of 0.6426, which are 3.28\% and 0.0649 higher than the second-best method, respectively.
Especially for the subject "ay" with only 28 training samples, our method still achieves a recognition accuracy of 67.86\%. This experimental result demonstrates the superior generalization ability of our method in small-sample EEG decoding tasks.

\begin{table}[t]
\centering
\begin{threeparttable}
\caption{Classification accuracy(\%) and kappa comparisons with SOTA methods on Dataset III.}
\label{dataset3}
\renewcommand{\arraystretch}{1.18}
\begin{tabularx}{0.95\textwidth}{lXXXXXXXX}
\hline
\multirow{2}{*}{Methods} & al      & aa      & av     & aw     & ay     & \multirow{2}{*}{Average} & \multirow{2}{*}{Std} & \multirow{2}{*}{Kappa} \\ \cline{2-6}
                         & 224/256 & 168/112 & 84/196 & 56/224 & 28/252 &                          &                        \\ \hline
EEGNet~\cite{lawhern2018eegnet}                  & 100     & 68.75   & 58.16  & 79.46  & 51.59  & 71.59 & 19.09                & 0.4331                 \\
FBMSNet~\cite{liu2022fbmsnet}                  & 100     & 82.14   & 57.14  & 82.04  & 59.33  & 76.13  & 17.91                  & 0.5270                 \\
LMDA-Net~\cite{miao2023lmda}                 & 100     & 70.13   & 61.33  & 78.94  & 52.11  & 72.50   & 18.33                & 0.4567                 \\
M-FANet~\cite{qin2024m}                  & 100     & 79.46   & 58.16  & 81.70   & 59.92  & 75.85  & 17.30                  & 0.5211                 \\
Basenet-SE~\cite{wimpff2024eeg}               & 100     & 79.46   & \underline{64.80}  & 75.89  & \underline{64.68}  & 76.97  & \underline{14.46}                  & 0.5374                 \\
ATCNet~\cite{altaheri2022physics}                   & 100     & 75.00   & 61.22  & 79.02  & 53.57  & 73.76  & 17.91                & 0.4750                 \\
Conformer~\cite{song2022eeg}                & 100     & \underline{83.93}   & 63.78  & \underline{82.14}  & 63.89  & \underline{78.75} & 15.29                   & \underline{0.5777}                 \\
MSVTNet~\cite{liu2024msvtnet}                & 100     & 74.11   & 63.78  & 81.70  & 62.70  & 76.46   & 15.31                 & 0.5354                \\
\textbf{\method}                   & \textbf{100}     & \textbf{88.39}   & \textbf{70.41}  & \textbf{83.48}  & \textbf{67.86}  & \textbf{82.03} & \textbf{13.24}                   & \textbf{0.6426}                 \\ 
\hline
\end{tabularx}
\begin{tablenotes}[flushleft] 
\item Best performances are highlighted in bold, while the second-best with underlined.
\end{tablenotes} 
\end{threeparttable}
\end{table}

\subsection{Ablation Study }

The significant improvement of our \method can be attributed to three novel designs: the Adaptive-Spatial-Spectral fusion module, the Multi-scale Variance Pooling module, and the Dual Prototype Learning approach. To further analyze the impact of these three modules on model performance, we conduct ablation experiments on Datasets I and II.
Four models, named Model1, Model2, Model3 and Model4, are utilized, which represent four scenarios as follows:
\begin{itemize}
    \item \textbf{Model1}
    The model is realized by removing the ASSF module and adopting a depthwise convolution used in EEGNet~\cite{lawhern2018eegnet} to fuse the information between EEG electrodes.
    \item \textbf{Model2}
    This model removes the MVP module and uses the separable convolution from EEGNet~\cite{lawhern2018eegnet} to extract temporal features.
    \item \textbf{Model3}
    This model removes the DPL module and uses CE loss to optimize parameters.
    \item \textbf{Model4}
    This model removes the DPL module and uses the PL method~\cite{yang2020convolutional} to optimize parameters.
\end{itemize}

\begin{table}[t]
\centering
\begin{threeparttable}
\caption{Classification accuracy (\%) and kappa for ablation experiments on Dataset I and Dataset II.}
\label{ablation}
\renewcommand{\arraystretch}{1.18}
\begin{tabular}{lccccccccccc}
\hline
\multirow{2}{*}{Model} & \multicolumn{5}{c}{Components} & \multicolumn{3}{c}{Dataset I} & \multicolumn{3}{c}{Dataset II} \\ \cline{2-12} 
                       & ASSF  & MVP  & CE  & PL  & DPL & Average   & Std     & Kappa   & Average    & Std    & Kappa    \\ \hline
Model1                 &       & \checkmark    &     &     & \checkmark   & 79.28     & 11.03   & 0.7328  & 84.65      & 9.72   & 0.6931   \\
Model2                 & \checkmark     &      &     &     & \checkmark   & 78.59     & 9.79    & 0.7145  & 84.28      & \textbf{9.06}   & 0.6857   \\
Model3                 & \checkmark     & \checkmark    & \checkmark   &     &     & 75.85     & 10.42   & 0.6780  & 83.83      & 9.66   & 0.6765   \\
Model4                 & \checkmark     & \checkmark    &     & \checkmark   &     & 79.13     & 9.44    & 0.7217  & 85.80      & 9.36   & 0.7161   \\
SST-DPN                & \checkmark     & \checkmark    &     &     & \checkmark   & \textbf{84.11}     & \textbf{8.30}    & \textbf{0.7881}  & \textbf{86.65}      & 9.10   & \textbf{0.7330}   \\ \hline
\end{tabular}
\end{threeparttable}
\end{table}

\begin{figure}[th]
  \centering
  \includegraphics[width=0.95\columnwidth]{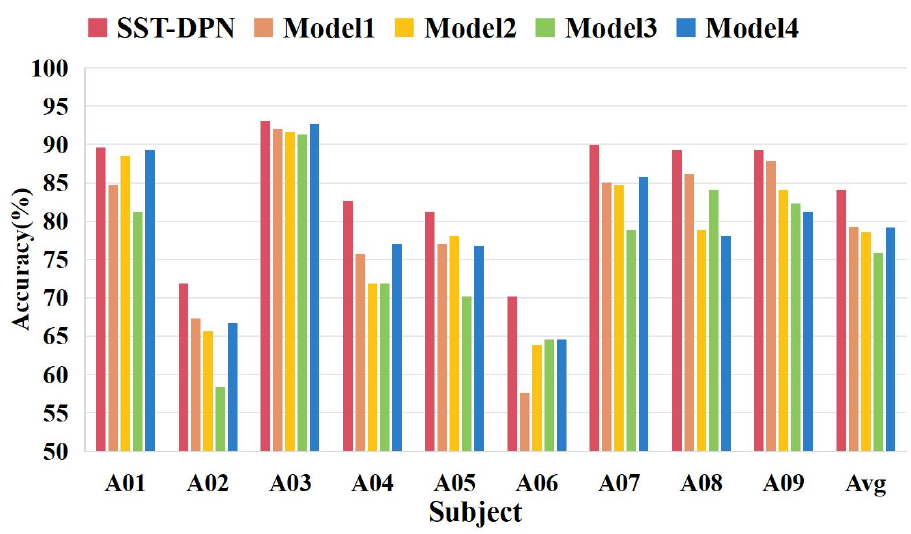}
  \caption{Accuracy ablation results for each subject on Dataset I. "Avg" indicates average accuracy.}
  \label{ablation1}
\end{figure}

\begin{figure}[!th]
  \centering
  \includegraphics[width=0.95\columnwidth]{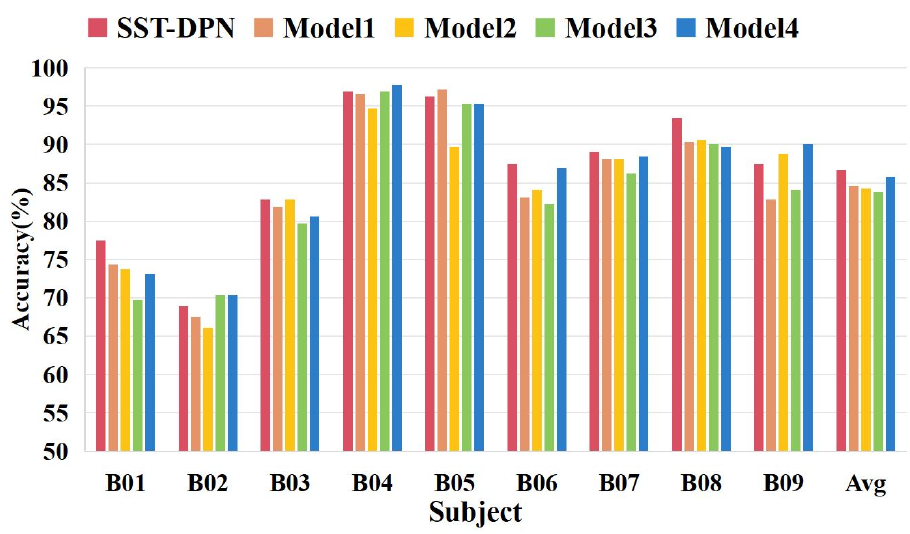}
  \caption{Accuracy ablation results for each subject on Dataset II. "Avg" indicates average accuracy.}
  \label{ablation2}
\end{figure}

\tablename~\ref{ablation} presents the ablation results for accuracy and kappa on Datasets I and II.
We also present detailed ablation results for each subject’s accuracy in \figurename~\ref{ablation1} and \figurename~\ref{ablation2}.
We first analyze the experimental results on Dataset I.
It can be clearly seen that the ASSF module brings a 4.83\% average accuracy improvement for Model1. 
This is because our ASSF module enables unified modeling of spatial-spectral features, and our use of SSA explicitly highlights the important spatial-spectral features.
Similarly, our MVP module effectively captures long-range temporal features, significantly enhancing the model's discriminative capability. As a result, our \method achieves an accuracy improvement of 5.52\% over Model2.
Employing dual prototype learning yields the most substantial enhancements.  Compared with Model3, it not only results in an 8.26\% average accuracy improvement but also results in consistent improvement on each subject. Our DPL approach, when compared to CE loss, not only increases intra-class compactness but also further enlarges inter-class margins. This greatly enhances the model's generalization capability and recognition performance.
Moreover, our \method has improved the accuracy by 4.98\% compared to Model4. This demonstrates that our DPL method has effectively improved upon the classical PL methods.
A similar result is also observed on Dataset II. As shown in \tablename~\ref{ablation} and \figurename~\ref{ablation2}, on Dataset II, our \method achieves accuracy improvements of 2.00\%, 2.37\%, 2.82\%, and 0.85\% compared to Model1, Model2, Model3, and Model4, respectively.

\begin{figure*}[!b]
\centering
\subfigure[Dataset I]{
\includegraphics[width=2.1in]{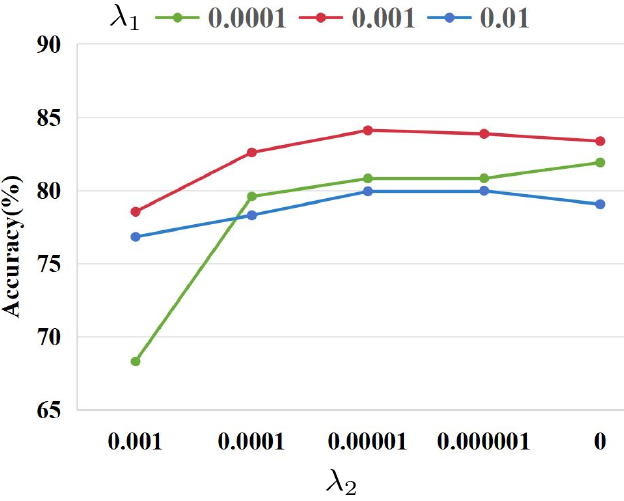}
}
\label{fig7:a}
\hspace{-0.4cm}
\subfigure[Dataset II]{
\includegraphics[width=2.1in]{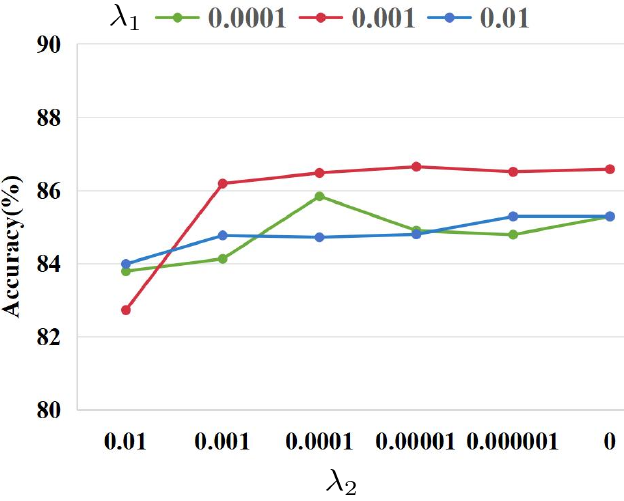}
}
\label{fig7:b}
\hspace{-0.4cm}
\subfigure[Dataset III]{
\includegraphics[width=2.1in]{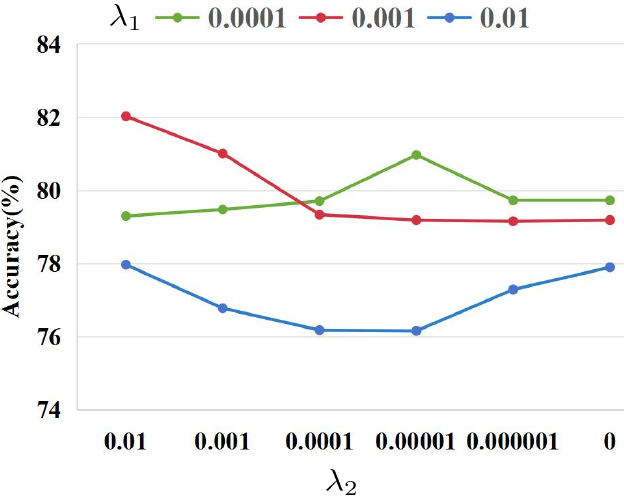}
}
\label{fig7:c}
\caption{The accuracy of \method across various settings of ${\lambda}_1$  and ${\lambda}_2$ on three datasets.}
\label{fig7}
\end{figure*}
\subsection{Parameter Sensitivity }
Our \method employs a combination of $\mathcal{L}_s$, $\mathcal{L}_c$ and $\mathcal{L}_{EF}$ as the final loss function, as shown in Eq.~\eqref{eq15}. While the $\mathcal{L}_s$ loss aims to minimize misclassification of subject movement intent, the $\mathcal{L}_c$ loss minimizes the sum of the embedded space distance of samples in a class to its center, making the samples belonging to the same class compact in the feature space.
And $\mathcal{L}_{EF}$ provides an explicit force, pushing the features away from the origin of the feature space to achieve larger inter-class margins.
The ${\lambda}_1$ and ${\lambda}_2$ are used to balance the impact of these three losses. 
To evaluate the influence of the ${\lambda}_1$ and ${\lambda}_2$, an empirical investigation compares the performance of \method across various settings on all three datasets.

As shown in \figurename~\ref{fig7}, ${\lambda}_1 = 0.001$ is the most suitable value across all three datasets. 
When ${\lambda}_1$ is increased to 0.01 or decreased to 0.0001, there is a noticeable decrease in accuracy on Datasets I and II. 
When ${\lambda}_1$ is fixed at 0.001 and ${\lambda}_2$ varies between 0 and 0.001,  the accuracies are consistently high and reach the best performance at ${\lambda}_2=0.00001$ on Dataset I and II. 
It is worth noting that even when ${\lambda}_2 = 0$, the accuracies on Datasets I and II remain high. 
This indicates that our DPL can automatically learn larger inter-class margins relying solely on the implicit force.
In contrast, on Dataset III, the best performance is achieved when ${\lambda}_2$ is increased to 0.01.
This implies that increasing the $\mathcal{L}_{EF}$ can further improve classification accuracy on datasets with fewer training samples.

In summary, our \method is relatively robust to the values of the hyperparameters \({\lambda}_1\) and \({\lambda}_2\). When \({\lambda}_1 = 0.001\) and \({\lambda}_2\) is set to a small value (i.e., \({\lambda}_2 < 0.001\)), \method can achieve good performance. If the amount of training data is very limited, further increasing \({\lambda}_2\) can be considered.

\section{Interpretable Analysis and Discussion}

\subsection{Effect of Adaptive Spatial-Spectral Fusion }
The primary insight of our ASSF module is to extract the discriminative features from multiple EEG electrodes within a unified spatial-spectral dimension. Unlike the classical EEGNet~\cite{lawhern2018eegnet}, which treats spatial and spectral features as separate dimensions, our ASSF module enables a more comprehensive and fine-grained modeling of spatial-spectral relationships. Additionally, we designed a lightweight SSA mechanism to emphasize important channels and suppress irrelevant ones within the spatial-spectral dimension, allowing the model to focus more effectively on features relevant to the current task.

\begin{figure}[!t]
\centering
\subfigure[A03 in Dataset I]{
\includegraphics[width=2.5in]{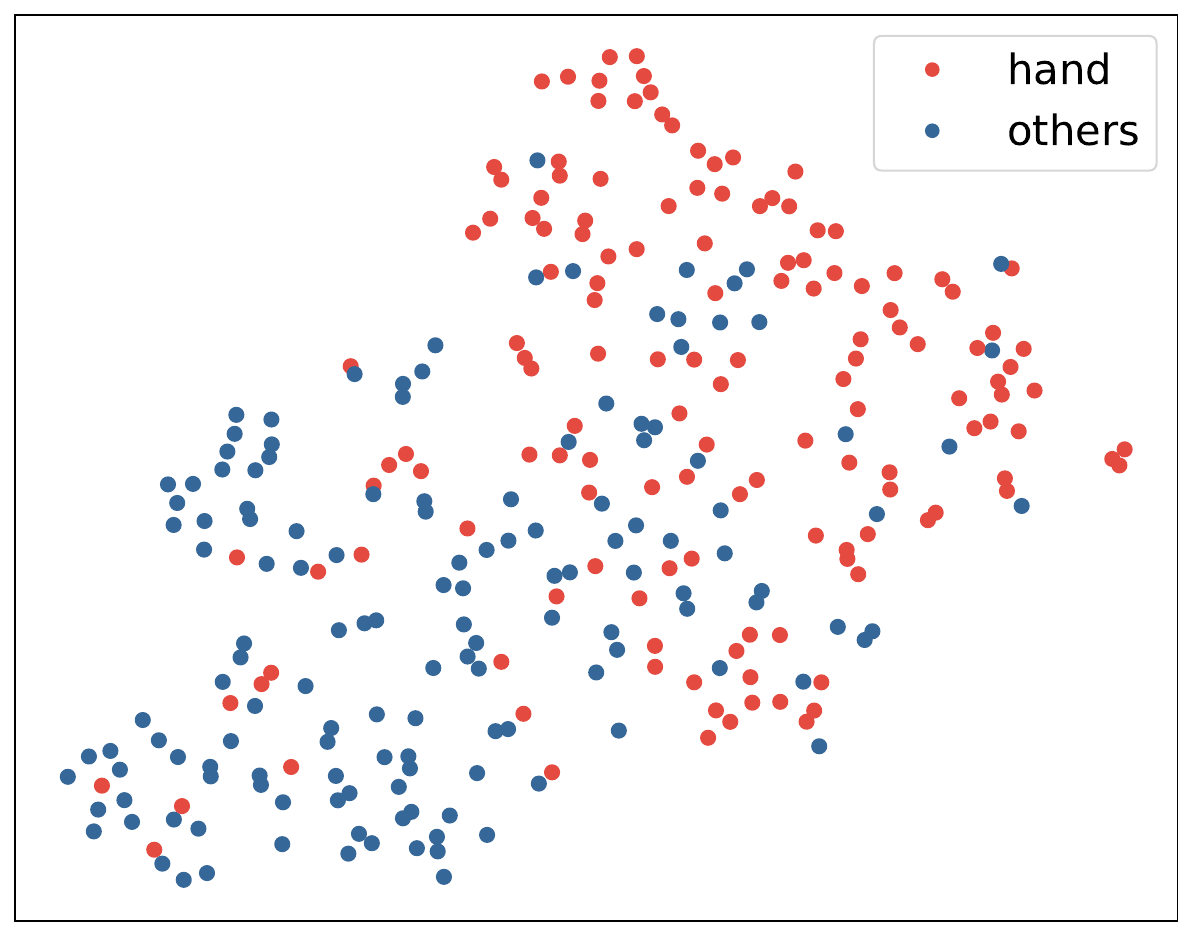}\label{ssa:a}
}
\hspace{0.3cm}
\subfigure["al" in Dataset III]{
\includegraphics[width=2.5in]{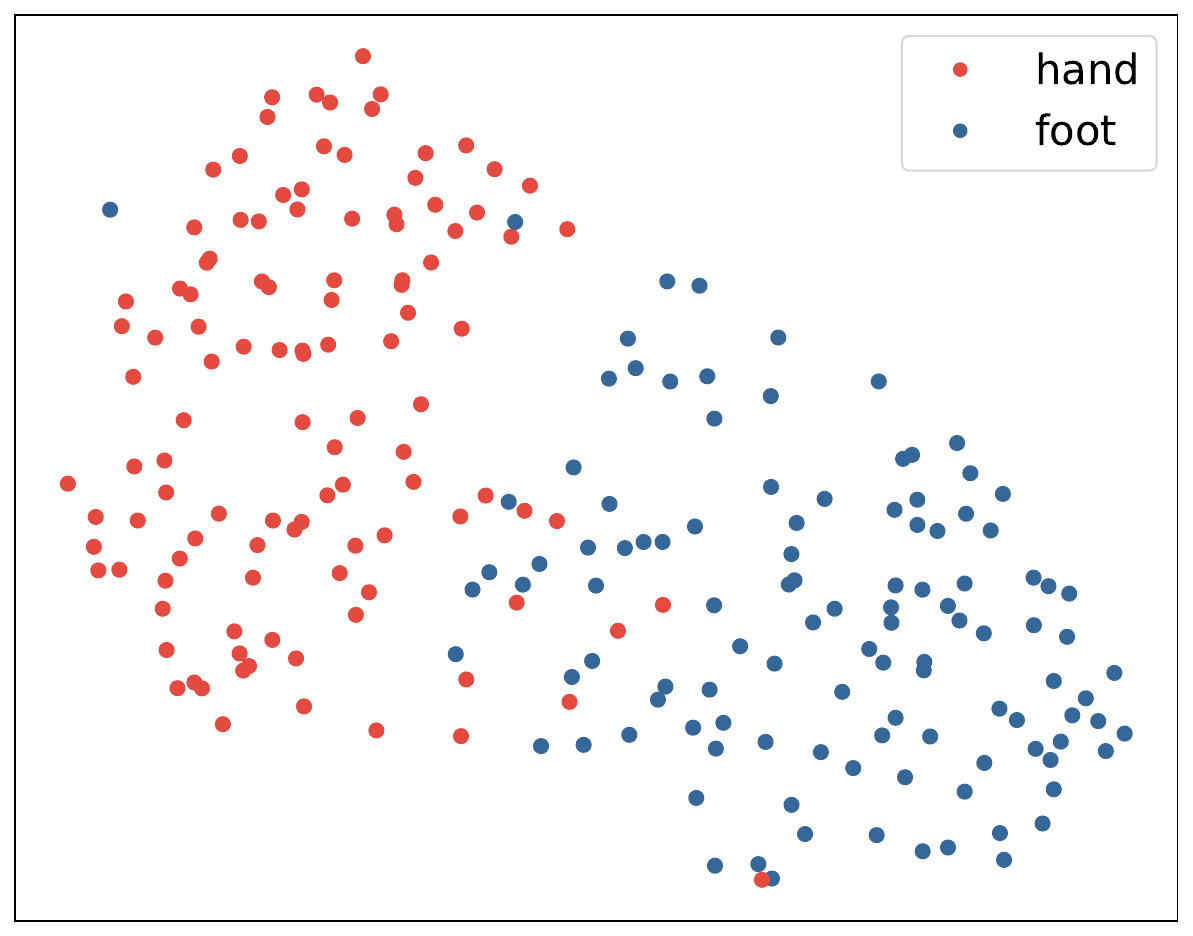}\label{ssa:b}
}
\caption{The distribution of attention vectors when performing different MI tasks on two datasets. All attention vectors are mapped to the 2D space using the t-SNE method. } 
\vspace{-0.36 cm}
\label{ssa}
\end{figure}

To verify the role of the SSA mechanism in imagining movements of different body parts, we visualize the attention vectors in Eq.~\eqref{eq6} using the t-SNE~\cite{van2008visualizing} method. 
As shown in \figurename~\ref{ssa}, there are distinct differences in the distribution of attention vectors when A03 performs MI of the hand and MI of others on Dataset I. Similarly, when the subject "al" in Dataset III
performs MI tasks of the hand and foot, the distribution of attention vectors also shows clear boundaries and clusters.
This provides strong evidence that the SSA mechanism can capture spatial-spectral channels crucial for the current MI task and that there are noticeable differences in the important channels across different MI tasks.

\begin{table*}[bth]
\centering
\caption{Comparison of classification accuracy (\%) and kappa between MVP and three other commonly used pooling layers in deep learning, as well as large-kernel depthwise convolution (LKDC), on Dataset I. 
All four comparison methods maintain the same multi-scale large-kernel design as MVP. LKDC is followed by necessary BatchNorm and ELU activation layers, while LpPooling uses the L2 norm.}
\label{lkdc}
\renewcommand{\arraystretch}{1.20}
\begin{tabularx}{\textwidth}{l *{10}{X}}  
\toprule
\multirow{2}{*}{Subjects} & \multicolumn{2}{c}{LKDC} & \multicolumn{2}{c}{MaxPooling} & \multicolumn{2}{c}{AvgPooling} & \multicolumn{2}{c}{LpPooling} & \multicolumn{2}{c}{MVP} \\ 
\cmidrule(lr){2-11} 
 & ACC & Kappa & ACC & Kappa & ACC & Kappa & ACC & Kappa & ACC & Kappa \\ 
\midrule
A01 & 56.94 & 0.4259 & 87.50 & 0.8333 & 87.15 & 0.8287 & 87.85 & 0.8380 & 89.58 & 0.8611 \\
A02 & 39.58 & 0.1944 & 62.15 & 0.4954 & 65.97 & 0.5463 & 62.15 & 0.4954 & 71.88 & 0.6250 \\
A03 & 76.04 & 0.6806 & 92.36 & 0.8981 & 94.79 & 0.9306 & 95.49 & 0.9398 & 93.06 & 0.9074 \\
A04 & 47.22 & 0.2963 & 70.49 & 0.6065 & 77.43 & 0.6991 & 76.74 & 0.6898 & 82.64 & 0.7685 \\
A05 & 34.38 & 0.1250 & 74.31 & 0.6574 & 75.00 & 0.6667 & 74.31 & 0.6574 & 81.25 & 0.7500 \\
A06 & 40.62 & 0.2083 & 68.40 & 0.5787 & 69.44 & 0.5926 & 69.79 & 0.5972 & 70.14 & 0.6019 \\
A07 & 62.50 & 0.5000 & 88.19 & 0.8426 & 89.24 & 0.8565 & 88.19 & 0.8426 & 89.93 & 0.8657 \\
A08 & 64.93 & 0.5324 & 83.68 & 0.7824 & 80.90 & 0.7454 & 87.50 & 0.8333 & 89.24 & 0.8565 \\
A09 & 63.54 & 0.5139 & 86.81 & 0.8241 & 80.21 & 0.7361 & 86.11 & 0.8148 & 89.24 & 0.8565 \\
\midrule
Average & 53.97 & 0.3863 & 79.32 & 0.7243 & 80.01 & 0.7336 & 80.90 & 0.7454 & 84.11 & 0.7881 \\
\bottomrule
\end{tabularx}
\end{table*}

\subsection{Effect of Multi-scale Variance Pooling }
Inspired by related work on large-kernel convolutions, our MVP module innovatively employs a pooling layer with large kernels to extract long-term temporal information.
Moreover, utilizing the crucial prior knowledge of spectral power in EEG signals, we design a variance pooling layer. 
Here, we discuss in greater detail the advantages of MVP in EEG-MI decoding. As shown in \tablename~\ref{lkdc}, we compare MVP with the large-kernel convolution method and three commonly used pooling layers in deep learning.
Consistent with conclusions in computer vision, large-kernel convolutions, though theoretically capable of capturing long-range dependencies, present training challenges. Without additional, carefully crafted training strategies, large-kernel convolutions can even reduce model performance. This challenge is especially notable in small-sample EEG-MI recognition tasks, where large-kernel convolutions are highly susceptible to overfitting.
In contrast, the other three large-kernel pooling layers achieved superior performance, surpassing that of the transformer models in \tablename~\ref{dataset1}. This strongly demonstrates the great potential of large-kernel pooling layers for extracting effective temporal features in EEG signals.
Furthermore, benefiting from variance pooling specifically designed for EEG signals, our MVP method further improves recognition performance compared to commonly used pooling layers in deep learning.

\begin{figure}[!b]
  \centering
    \vspace{-0.36cm}
  \includegraphics[width=0.88\columnwidth]{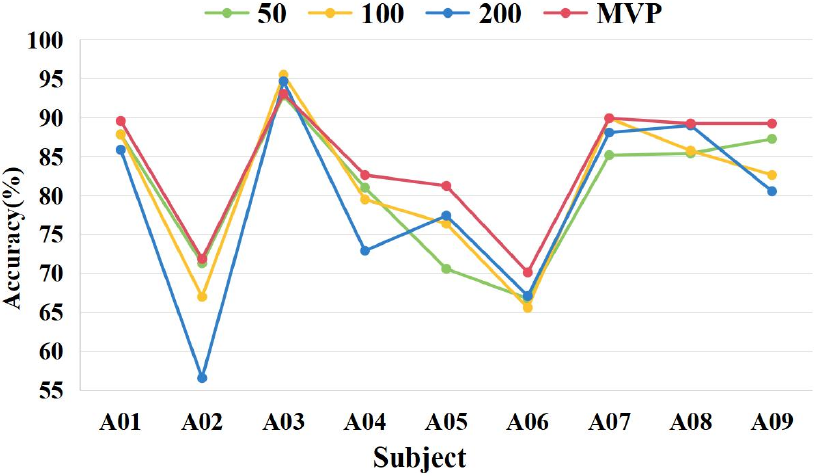}
    \vspace{-0.2cm}
  \caption{Comparison of accuracy between single-scale variance pooling with different kernel sizes and MVP. }
  \vspace{-0.36cm}
  \label{mvp}
\end{figure}

Next, we explore the role of the multi-scale design in the MVP module.
Within one EEG trial, the start point and duration of the actual MI period showing the appropriate ERS and ERD pattern can be different from trial to trial~\cite{deny2023hierarchical}.
This phenomenon is more significant among trials between different subjects. 
In order to adapt to these differences between trials and extract more discriminative temporal information, we group the EEG representations along the channel dimension and use variance pooling with different kernel sizes to extract multi-scale temporal information. 
As shown in \figurename~\ref{mvp}, a smaller kernel size of 50 performs better on subjects A02, A04, and A09, while a larger kernel size of 200 performs better on subjects A05, A07, and A08. 
Altogether, the MVP integrates information from different scales and achieves the best overall performance across all subjects.

\begin{figure*}[!t]
\centering
\subfigure[CE loss]{
\includegraphics[width=3.15in]{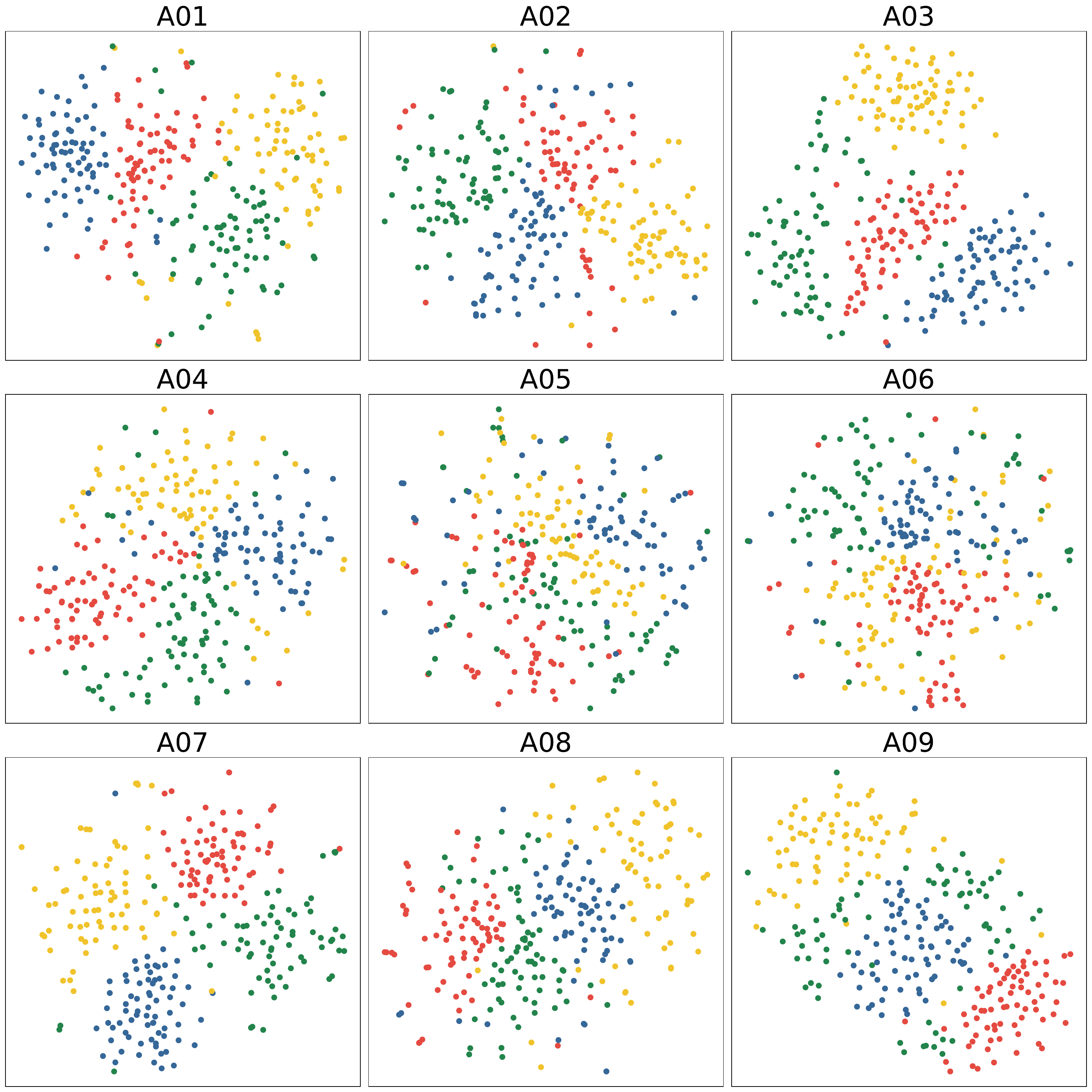}
}
\label{fig10:a}
\subfigure[DPL]{
\includegraphics[width=3.15in]{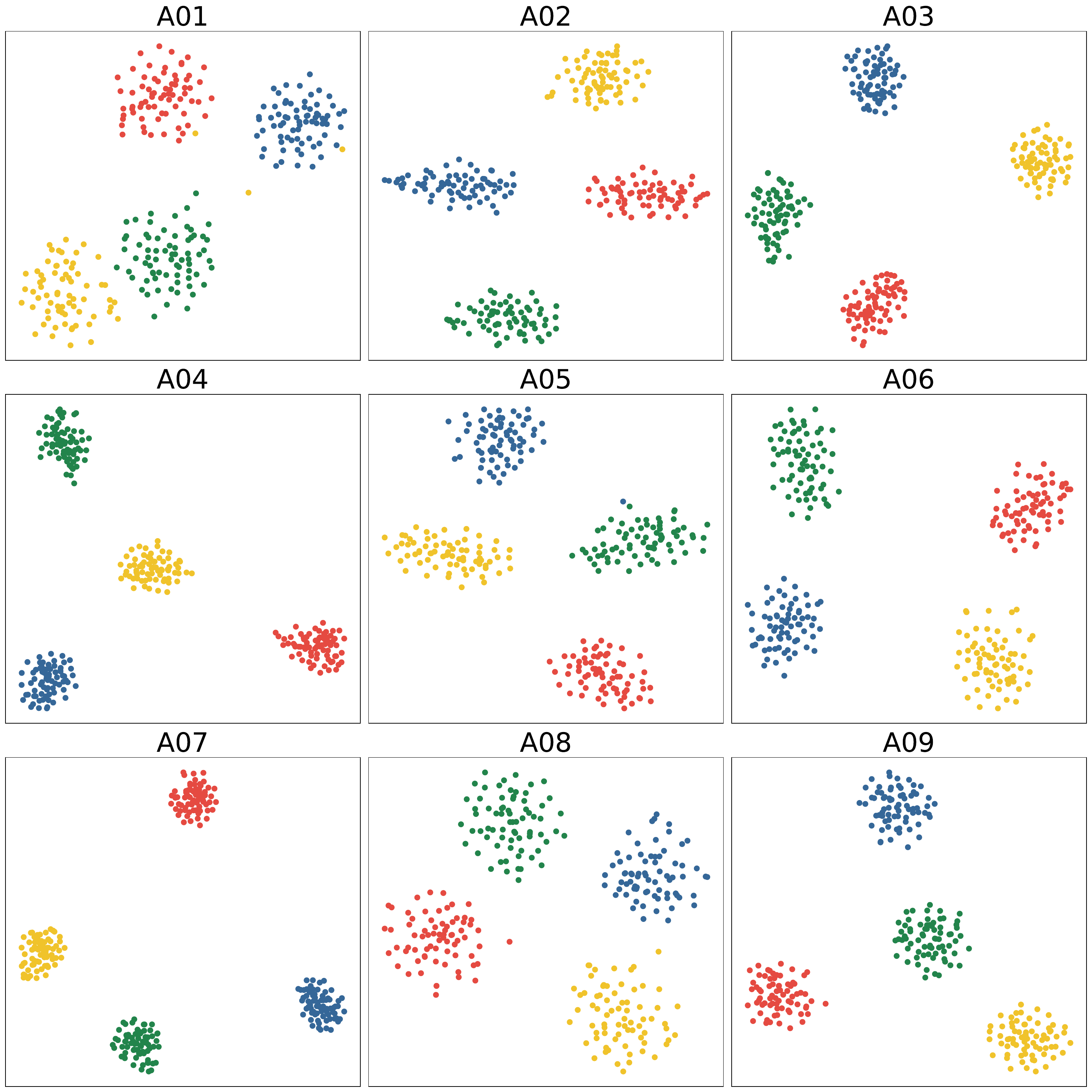}
}
\label{fig10:b}
\caption{t-SNE visualization of the feature distribution for each subject trained on Dataset I using CE loss and DPL. The points with different colors denote features from different classes.}
\label{fig10}
\end{figure*}

\begin{figure}[t]
\centering
\subfigure[Accuracy curve on the training set]{
\includegraphics[width=3.3in]{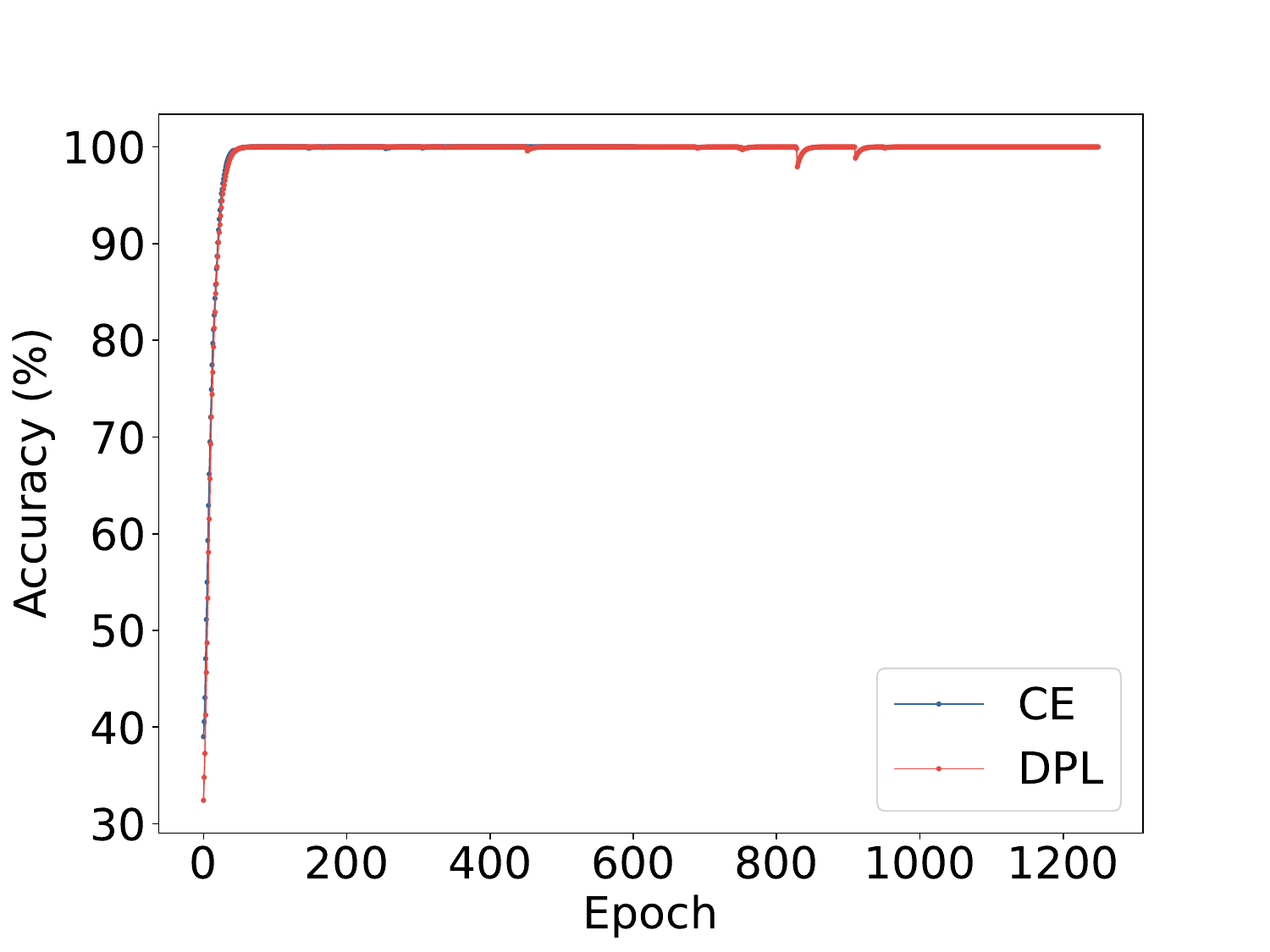}\label{acc:a}
}
\hspace{-0.9cm}
\subfigure[Accuracy curve on the test set]{
\includegraphics[width=3.3in]{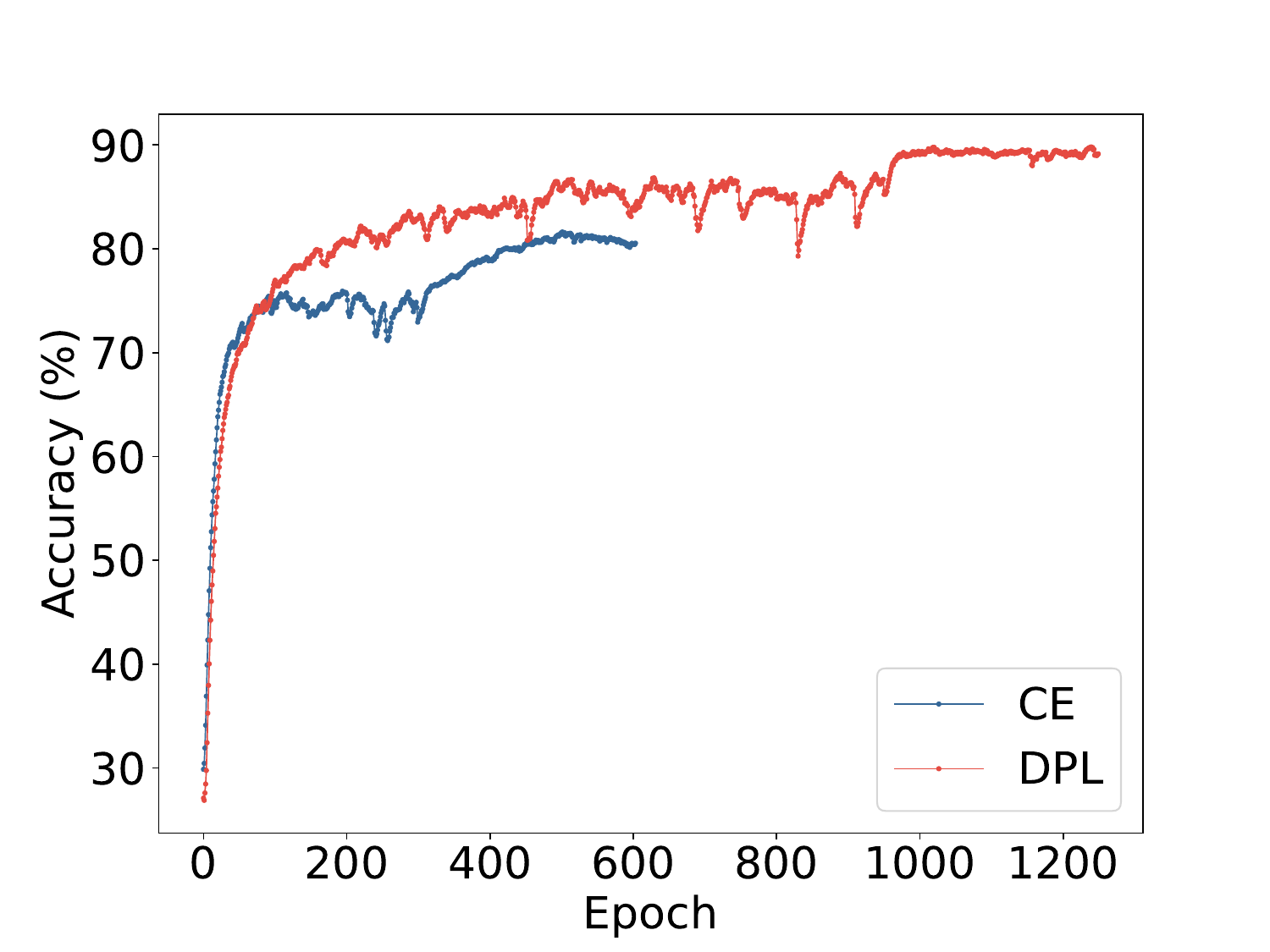}\label{acc:b}
}
\caption{Accuracy curve changes on the training and test sets for subject A01 on Dataset I when trained with CE and DPL. Please note that in (a), the accuracy curves of CE and DPL on the training set are very close; zooming in can provide a clearer view. Since we used an early stopping mechanism on the validation set in the first training stage, the total training epochs differ between the two training methods.} 
\label{acc}
\end{figure}

\subsection{Effect of Dual Prototype Learning }
Our \method is the first to apply prototype learning methods to MI-EEG decoding. Furthermore, we decouple inter-class separation and intra-class compactness by using two prototypes, ISP and ICP, for each class. \figurename~\ref{ablation1}, \figurename~\ref{ablation2}, and \tablename~\ref{ablation} demonstrate that our DPL method significantly improves the recognition accuracy of MI tasks compared to classic CE loss. 
To further explain the effectiveness of DPL, we visualize the distribution of feature vectors $z$ using the t-SNE method. 
\figurename~\ref{fig10} displays the feature distribution of all subjects on Dataset I under both CE loss and DPL optimization approaches.
It is apparent that our DPL method achieves greater intra-class compactness and larger inter-class margins compared to CE loss.
Therefore, our DPL significantly enhances the model's generalization ability.
This also intuitively explains why our DPL achieves better classification accuracy.

We further examine the accuracy curve changes for CE and DPL on both the training and test sets over successive training epochs. Specifically, we select A01 from Dataset I, testing accuracy on the test set after each training epoch\footnote{This experiment is solely intended to clarify how DPL effectively prevents overfitting compared to CE. In our main experiments, the test set remains strictly unseen during training, as described in Section 4.2.1}. As shown in \figurename~\ref{acc}, both CE and DPL quickly fit the training data, reaching 100\% accuracy. However, CE exhibits notably lower accuracy on the test set compared to DPL, indicating overfitting on the training set. In contrast, our DPL method optimizes the feature space distribution, effectively reducing the risk of overfitting and maintaining strong performance on the test set.

\begin{figure*}[h]
  \centering
  \includegraphics[width=1.0\textwidth]{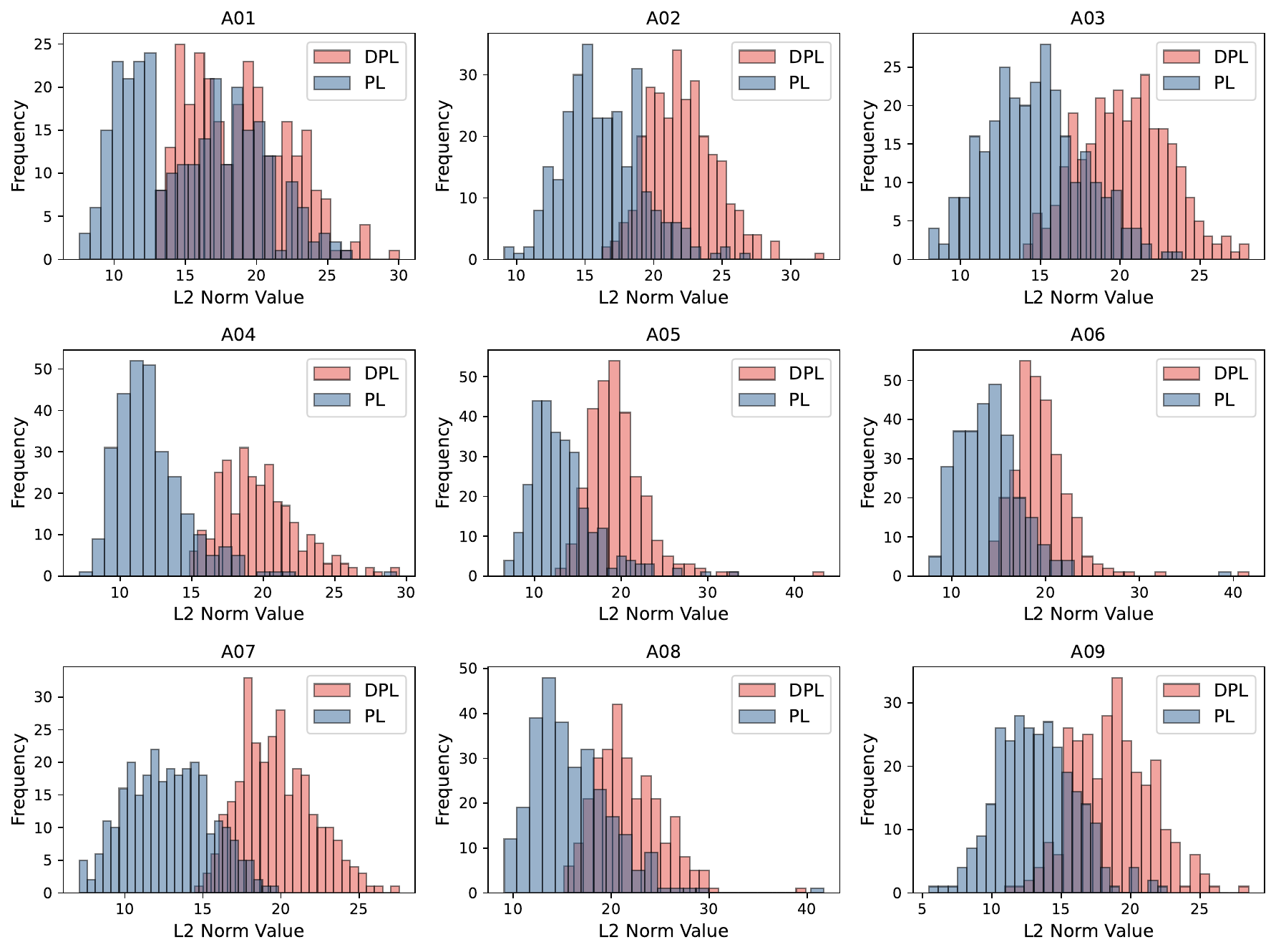}
  \caption{Histogram of the L2 norm distribution of features for each subject trained on Dataset I using PL and DPL, respectively. }
  \label{fig11}
\end{figure*}

To verify that our DPL pushes the features further away compared to the classic PL method, thereby achieving larger inter-class margins, we visualize the feature norm distribution of our DPL method and the PL method. 
\figurename~\ref{fig11} shows the L2 norm distribution of deep features trained with DPL and PL on Dataset I.
It can be clearly seen that across all subjects, the feature norms trained using DPL are statistically larger than those trained using PL. 
This realizes the feature space optimization process from \figurename~\ref{fig4} (b) to \figurename\ref{fig4} (d), demonstrating the superiority of DPL.

\subsection{Computational Costs }
As BCI systems typically operate in online or closed-loop mode on devices with limited computational resources~\cite{wimpff2024eeg,huang2024ltdnet}, it is crucial to examine the computational expense of new algorithms.
\tablename~\ref{costs} displays the preprocessing methods, the classification accuracy, model parameters, and Floating Point Operations (FLOPs) for Dataset I. 
LMDA-Net needs to use Euclidean alignment (EA) to achieve high recognition performance, but EA is not suitable for real-time testing.
FMBSNet and M-FANet require time-consuming multi-narrow-band band-pass filtering. 
In comparison, our \method does not require any preprocessing steps and achieves the highest recognition accuracy with the lowest FLOPs. 
The three transformer models have a significantly higher parameter count compared to other methods.
This suggests that our model optimally balances the accuracy and speed of MI-EEG decoding. Moreover, the number of parameters for our model is much less than transformer-based methods and is on par with lightweight CNN methods.

\begin{table}[!ht]
\centering
\begin{threeparttable}
\caption{Comparisons with SOTA methods in the computational costs and recognition performance on Dataset I.}
\label{costs}
\renewcommand{\arraystretch}{1.18}
\begin{tabularx}{0.95\textwidth}{llXXXX}
\hline
Methods   & Preprocessing    & Acc (\%) & Kappa  & Parameters (k) & FLOPs (M) \\ \hline
LDMA-Net~\cite{miao2023lmda}  & EA \& BP         & 75.40   & 0.6700 & \textbf{3.71}          & 50.38    \\
M-FANet~\cite{qin2024m}   & MBP              & 79.28   & 0.7259 & 4.08          & 23.39    \\
Basenet-SE~\cite{wimpff2024eeg}   & BP              & 76.00   & 0.6794 & 3.82          & 26.64    \\
FBMSNet~\cite{liu2022fbmsnet}   & MBP              & 79.17   & 0.7235 & 16.23         & 99.95    \\
Conformer~\cite{song2022eeg} & BP               & 75.27   & 0.6702 & 789.57        & 63.86    \\
ATCNet~\cite{altaheri2022physics}    & no preprocessing & 79.86   & 0.7312 & 113.73        & 29.81    \\
MSVTNet~\cite{liu2024msvtnet}    & no preprocessing & 75.50   & 0.6733 & 75.49        & 51.56    \\
\textbf{\method}      & no preprocessing & \textbf{84.11}   & \textbf{0.7881} & 15.21         & \textbf{9.65}     \\ \hline
\end{tabularx}
\begin{tablenotes}[flushleft]
\item BP: band-pass filtering, MBP: multi-narrow-band band-pass filtering.
\end{tablenotes}
\end{threeparttable}
\end{table}

\subsection{Limitation and Future Work}
Although the proposed \method addresses major challenges in EEG-MI decoding and achieves excellent performance, there are some limitations in our current work. 
First, incorporating more prior knowledge into neural network design is worth exploring, such as considering the mirror distribution of EEG electrodes and the functional partitioning of the brain~\cite{zhang2023local,luo2023shallow}.
Although the LightConv proposed in the ASSF module could potentially leverage this prior knowledge, we did not further explore this aspect as it is not the focus of this work. 
Second, the potential of the brain-inspired DPL framework remains to be fully explored. 
By decoupling inter-class separation and intra-class compactness, we simply constrain the prototype and feature distribution to achieve superior performance. 
Nonetheless, more effective and discriminative loss functions deserve further investigation. 
Finally, the \method has only undergone offline testing on public datasets and has not yet been validated in an online BCI environment. 
In the future, we will continue to enhance \method based on these avenues to achieve good performance in online BCI applications.

\section{Conclusion}
\label{sec:Conclusion-outlook}
In this paper, we propose a novel end-to-end framework for MI-EEG decoding.
SST-DPN is a unified learning framework that encompasses three key aspects of EEG-MI decoding: multi-channel spatial-spectral features, long-term temporal features, and the small-sample dilemma.
First, we extract multiple spectral features from each EEG electrode to form a spatial-spectral representation, followed by a carefully designed attention mechanism to explicitly model the spatial-spectral relationships across multiple channels. 
This facilitates the extraction of more robust and task-relevant spatial-spectral features, a critical aspect often underexplored in current research.
Subsequently, we propose a novel MVP module to capture long-term temporal features. With its low computational cost and ease of training, the MVP module presents a promising alternative to transformers for capturing global dependencies in EEG signal decoding.
Moreover, we introduce a novel DPL approach to explicitly increase intra-class compactness and inter-class margins in the feature space. 
The DPL enhances the model's generalization capability, thereby helping to alleviate the limited sample issue.
We conduct extensive experiments on three public datasets, and the results confirm that our method surpasses other SOTA methods.
Additionally, we conduct a detailed analysis and discussion of our results from the perspective of feature visualization, providing neurophysiological interpretations.
Together, the proposed \method holds promising potential for practical MI-BCI applications due to its remarkable performance combined with low computational costs.

\section*{Declaration of competing interest}
The authors declare that they have no known competing financial interests or personal relationships that could have appeared to influence the work reported in this paper.

\section*{Declaration of generative AI and AI-assisted technologies in the writing process}
During the preparation of this work the authors used ChatGPT in order to improve language and readability. After using this tool/service, the authors reviewed and edited the content as needed and take full responsibility for the content of the publication.

\section*{Acknowledgments}
This work was partially supported by OYMotion Technologies. 

\bibliographystyle{elsarticle-num}
\bibliography{refs}

\end{document}